\documentclass[fleqn,usenatbib]{mnras}
\usepackage{etoolbox}
\robustify\cite
\usepackage{xcolor}

\usepackage{newtxtext,newtxmath}

\usepackage[T1]{fontenc}
\usepackage[countmax]{subfloat}

\DeclareRobustCommand{\VAN}[3]{#2}
\let\VANthebibliography\thebibliography
\def\thebibliography{\DeclareRobustCommand{\VAN}[3]{##3}\VANthebibliography}

\usepackage{graphicx} 
\usepackage{amsmath} 
\usepackage{natbib}
\usepackage{enumitem}
\usepackage{booktabs}
\usepackage{pdflscape}
\usepackage{lscape}
\usepackage{longtable}

\usepackage{booktabs}
\usepackage{amsmath}
\usepackage{array}
\usepackage{wasysym}
\newcommand{\msun}{M$_\odot$}

\newcommand{\msy}{M$_\odot$~yr$^{-1}$}

\newcommand{\mbh}{M$_\bullet$}
\newcommand{\mmbh}{M_\bullet}

\newcommand{\LCDM}{$\Lambda$CDM}

\newcommand{\Htwo}{H$_{2}$}

\newcommand{\HeI}{\ion{He}{i}}

\newcommand{\Fe}{[\ion{Fe}{ii}]}

\newcommand{\SII}{[\ion{S}{ii}]}

\newcommand{\OIII}{[\ion{O}{iii}]}

\newcommand{\NII}{[\ion{N}{ii}]}

\newcommand{\SiI}{\ion{Si}{i}}

\newcommand{\MgI}{\ion{Mg}{i}}
\newcommand{\FeI}{\ion{Fe}{i}}
\newcommand{\AlI}{\ion{Al}{i}}
\newcommand{\CI}{\ion{C}{i}}
\newcommand{\CaI}{\ion{Ca}{i}}
\newcommand{\NaI}{\ion{Na}{i}}

\newcommand{\Ha}{H\,$\alpha${}}
\newcommand{\Hb}{H\,$\beta${}}

\newcommand{\pab}{Pa\,$\beta${}}

\newcommand{\pag}{Pa\,$\gamma${}}
\newcommand{\brg}{Br\,$\gamma${}}
\newcommand{\brd}{Br\,$\delta${}}

\newcommand{\ecs}{erg~cm\pwr{-2}~s\pwr{-1}}
\newcommand{\es}{erg~s\pwr{-1}}

\newcommand{\kms}{km~s\pwr{-1}}

\newcommand{\jk}{$J-K$}

\newcommand{\ebv}{$E(B-V)$}

\newcommand{\pwr}[1]{$^{#1}$}

\makeatletter
\newcommand*{\rom}[1]{\expandafter\@slowromancap\romannumeral #1@}
\makeatother
\makeatletter
\def\blfootnote{\xdef\@thefnmark{}\@footnotetext}
\makeatother

\newcolumntype{R}[1]{>{\RaggedLeft\arraybackslash}p{#1}}
\newcolumntype{C}[1]{>{\centering\arraybackslash}p{#1}}

\title[NIR Spectral Atlas of ETGs]{Infrared Spectroscopy of Nearby Radio Active Early-Type Galaxies - II: Spectral Atlas}

\author[M. Durr\'{e} et al.]{
Mark Durr\'e,$^{1}$\thanks{E-mail: mdurre@swin.edu.au}
Jeremy Mould,$^{1}$
Michael Brown$^{2}$ and
Tristan Reynolds$^{3,4}$
\\
$^{1}$Centre for Astrophysics \& Supercomputing, Swinburne University, Hawthorn VIC 3122, Australia\\
$^{2}$School of Physics, Monash University, Clayton, Vic 3800, Australia\\
$^{3}$International Centre for Radio Astronomy Research (ICRAR), The University of Western Australia, 35 Stirling Highway, Crawley, WA 6009, Australia\\
$^{4}$ARC Centre of Excellence for All Sky Astrophysics in 3 Dimensions (ASTRO 3D), Australia\\
}

\date{Accepted XXX. Received YYY; in original form ZZZ}

\pubyear{2023}
\defcitealias{Brown2011}{B11}
\defcitealias{Mould2012}{M12}
\begin{document}
\label{firstpage}
\pagerange{\pageref{firstpage}--\pageref{lastpage}}
\maketitle
\begin{abstract}
We present a near infrared spectroscopic atlas of nearby, bright early-type galaxies with radio emission, containing 163 galaxies observed by the Palomar 200\arcsec{} TripleSpec instrument, measuring the emission line fluxes for H, He, \Fe{} and \Htwo{} and determined the nuclear excitation mechanisms. By stacking spectra, we deduced the \Htwo{} excitation temperature ($1957\pm182$ K) and dominant excitation mechanism (thermal and shock heating combined) from the \textit{K}-band emission line sequence. Stacking also produces an ``average'' spectrum of absorption features and spectral indices from the literature; the CO12 absorption line index vs. \jk{} colour shows a trend of stronger nuclear activity producing a weaker CO12 index and a redder (flatter) continuum. The correlations between the radio and the emission-line luminosities finds a trend with radio power; however, the large scatter in the upper limits shows that the two are not directly coupled and the duty cycles of SF and AGN activity are not synchronised. 
\end{abstract}




\section{Introduction}
Early-type galaxies (ETGs) have several advantages for spectroscopic studies of nuclear activity; their fuelling and stellar population is likely to be the simplest, their stored gas is smaller, there is a minimum of nuclear obscuration and nuclear outflows and jets have less ISM to interact with.  Near infrared (NIR) observations allow for dust penetration, getting ``closer'' to the central source \citep{Kennicutt2009,Brown2011,Mould2012,Ricci2021}.

It is highly likely that all massive early-type galaxies (M$_K$<-25.5) are radio continuum sources, presumably harbouring active galactic nuclei (AGN) or undergoing star formation \citep[][hereafter B11]{Brown2011}. \citetalias{Brown2011} provided an homogeneous sample of 396 bright early-type galaxies (m$_K$ < 9, T type < -1, i.e. Hubble types E and S0), with radio continuum measurements from the NRAO Very Large Array Sky Survey, Green Bank 300 ft Telescope, and 64 m Parkes Radio Telescope; this catalog is well-suited to spectroscopic and multi-wavelength follow-up. This work republishes the complete early-type radio galaxy catalog from \citetalias{Brown2011}, supplemented with S0/a galaxies and objects that were observed as described in this paper.

\cite{Sabater2019} cross-compared the LOFAR Two-Metre Sky Survey (LoTSS) with the Sloan Digital Sky Survey (SDSS) DR7 main galaxy spectroscopic sample, finding the fraction of radio AGN reached 100\% at >10\pwr{11}~\msun. \cite{Capetti2022}, from their catalog of 188 giant ETGs ($M_K \leq -25$) in their LOFAR sample, obtained 44 SDSS spectra and classified them using \cite{Kewley2006} diagnostics. Of these, the optical spectrum diagnostics showed 11 were AGN (all LINERs) and 7 were star forming (SF); the SF galaxies had a star formation rate (SFR) in the range 0.1-8 \msy. The rest (26 galaxies) did not have all the diagnostic emission lines detected with enough confidence to be placed on excitation diagrams.

\cite{Mould2012} (hereafter M12) published the results of near-infrared long-slit spectroscopic observations of 136 galaxies from the \citetalias{Brown2011} sample. These were observed in 5 sessions; on the Hale 5m Palomar TripleSpec spectrograph \citep{Herter2008} (3 sessions) and on the Mayall 4m KPNO Flamingos spectrograph (2 sessions). Of these 20\% showed nuclear infrared emission lines, mostly Paschen $\beta$ and Brackett $\gamma$ hydrogen lines and \Fe{} lines. Subsequently, further observing sessions using TripleSpec extended the number of objects observed by 96. This work presents the results from the complete observation program and makes the spectra available online.

These observations enable examination of the following science:
\begin{enumerate}[leftmargin=*,align=left]
    \item The gaseous excitation mechanism (AGN, star formation or high line ratio- HLR), determined by the emission line flux ratios of H, \Fe{} and \Htwo.
    \item The SFR in the nuclear region (if the object is not an AGN).
    \item The warm molecular hydrogen temperature and excitation mechanisms; the \Htwo{} luminosity also estimates the total molecular mass in the nuclear region.
    \item Correlations between radio and NIR emissions to examine AGN duty cycles in feeding/feedback models.
    \item Absorption line studies to determine stellar populations, especially thermally pulsing asymptotic giant branches (TP-AGB) stars which should dominate in younger populations \citep{Martins2013}.
\end{enumerate}

In this work, we determine redshift-dependent distances using the \LCDM{} cosmology with parameters $H_0 = 70$ km{} s$^{-1}$ Mpc$^{-1}$, $\Omega_{m} = 0.3$ and $\Omega_{\Lambda} = 0.7$.

\section{Observations and Data Reduction}
\subsection{Observations}
The \citetalias{Brown2011} catalog was extended in a follow-up survey to galaxies with morphological T types >= 0 (i.e. Hubble types S0/a and S), increasing the total to 514 objects. This extended list was used as the basis for the \citetalias{Mould2012} observations. This was supplemented by inclusion of objects that were observed using TripleSpec as targets of opportunity, but were not included in the previous lists. In total, there are 546 objects in the complete catalog (396 in \citetalias{Brown2011}, 118 in the extended survey and 32 TripleSpec supplementary objects). We publish the complete radio survey in this work (see Section \ref{sec:TSpecRadioCatalog} below).

Eight observing runs were conducted with the Triplespec instrument subsequent to the data published in \citetalias{Mould2012}, making a total of 11 runs. A complete list of these runs are detailed in Table \ref{tbl:TSpecAtlas01}, under programs P16, P22, P23 and P220. For each night, we list the number of objects observed that are from the original \citetalias{Brown2011} catalog (Set 1 - 88 galaxies), the extended catalog (Set 2 - 43 galaxies) and the supplementary objects (Set 3 - 32 galaxies), a total of 163 galaxies. Table \ref{tbl:TSpecAtlas02} lists the observational data, showing the galaxy name and observational details. This table is a sample; the complete list is available online, as described below. The detailed attributes for each galaxy may be found in Table \ref{tbl:TSpecAtlas05}. 

In general, the galaxies were observed in cycles in the standard ABBA mode, separated by 20\arcsec{} in position on the slit. Each exposure was 300 s, for a total of 1200 s. 20 galaxies were observed with more than one cycle; these are listed in the table where, for example, 1.5 cycles is an observation of ABBAAB positions. The slit aperture was 1\arcsec, and the slit was oriented in the east-west direction on the sky.
\begin{table}
\centering
\caption{Observation runs with the Hale 5m Palomar TripleSpec spectrograph.}
\label{tbl:TSpecAtlas01}
\begin{tabular}{@{}ccrrrr@{}}
\toprule
Date (UT)  & Run ID & Set 1 & Set 2 & Set 3 & Total \\ \midrule
2011-09-15 & 1  & 17& 1 & 0  & 18 \\
2011-09-16 & 2  & 22& 0 & 0  & 22 \\
2012-01-03 & 3  & 15& 0 & 0  & 15 \\
2012-05-28 & 4  & 5 & 0 & 0  & 5  \\
2012-05-31 & 5  & 7 & 0 & 0  & 7  \\
2013-02-25 & 6  & 18& 2 & 2  & 22 \\
2013-07-31 & 7  & 0 & 0 & 15 & 15 \\
2013-08-17 & 8  & 0 & 0 & 15 & 15 \\
2013-12-14 & 9  & 0 & 14& 0  & 14 \\
2014-02-14 & 10 & 2 & 13& 0  & 15 \\
2014-02-15 & 11 & 2 & 13& 0  & 15 \\ \midrule
Totals  & & 88& 43& 32 & 163\\ \bottomrule
\end{tabular}
\end{table}

As noted in \citetalias{Mould2012}, the Flamingos observations produced poor results, due to lack of a good data reduction pipeline and instrument flexure. The instrument also does not cover the \textit{K}-band. However, it was useful for confirming the presence of emission lines (e.g. \pab{} and \Fe). These observations are not presented in this work. 
\begin{table*}
\renewcommand{\arraystretch}{1.1}
\centering
\caption{Sample object and observation details. (The full table is available online). See Table \ref{tbl:TSpecAtlas05} for the complete properties of each object.}
\label{tbl:TSpecAtlas02}
\begin{tabular}{@{}lrrrrclr@{}}
\toprule
Object&  RA& Dec& Cycles & Airmass & Run ID & Standard Star & S/N \\ 
\midrule
2MASX~J16251687-0910524& 246.3204& -9.1811& 1& 1.415& 8& HD~210290& 149.5\\
IC~0630& 159.6401& -7.1707& 1& 1.344& 3& HD~77281& 244.0\\
Mrk~612& 52.6703& -3.1377& 1& 1.425& 6& HD~21019& 88.8\\
NGC~~0016& 2.2678& 27.7294& 1& 1.15& 2& HD~221& 195.5\\
NGC~~1779& 76.3251& -9.1472& 1& 1.355& 11& HD~37887& 225.6\\
NGC~~2110& 88.0474& -7.4563& 1& 1.735& 2& HD~39439& 188.8\\
NGC~~2128& 91.1426& 57.6277& 1& 1.366& 1& HD~41654& 174.6\\
UGC~3024& 65.6108& 27.2979& 1& 1.211& 6& HD~283593& 130.9\\
UGC~3426& 93.9012& 71.0375& 1& 1.631& 1& HD~42507& 230.5\\
\bottomrule
\end{tabular}
\end{table*}
\subsection{Data Reduction}
The data reduction was carried out as described in \citetalias{Mould2012}, using the SpexTool v4.0 software \citep{Cushing2004}. The software pipeline creates normalized flats and wavelength calibrations (from OH telluric lines) for each night, then for each A--B pair applies flat field correction, traces object spectra,  defines extraction apertures, does background subtracting and extracts and wavelength calibrates spectra. As standard, the spectra are extracted in a window 1\arcsec.5 either side of the centre-line (i.e. 3\arcsec{} total width), and the background starts 3\arcsec{} from the centre-line. It then combines individual extracted (multi-order) spectra and performs telluric correction and flux calibration, using a standard A0 star which is matched to the spectrum of Vega. It then merges the orders into a single spectrum.

In post-processing on the output of SpexTool, we re-dispersed the original (non-linear) spectrum to a linear wavelength scale (1000--2440 nm) at 0.2 nm resolution (the average over the wavelength range). The TripleSpec instrument has a documented wavelength resolution $\Delta\lambda/\lambda = 2600$, which is equivalent to about 0.65 nm in the middle of the spectral range ($\sim$ 943--2465 nm). The spectra are cleaned by removing the wavelength ranges 1115--1150 nm, 1330--1480 nm and 1800--1950 nm; these are the atmospheric absorption bands between the \textit{Z}, \textit{J}, \textit{H} and \textit{K} windows. 

The spectra were then inspected and manually cleaned of noise spikes and increased noise near the absorption bands - this was more prominent on spectra with low flux. For a few spectra, the whole was manually cleaned, especially for those that had spectral features near the atmospheric windows (e.g. the \HeI{} 1083 nm line of NGC~~5548, which would otherwise lose the line wing in the gap). The spectra were then wavelength shifted to restframe. On inspection, almost half the spectra required removal of residual hydrogen line features; these were at zero redshift, caused by imperfect matching of the standard star to the Vega A0 model. These were removed by fitting single (or in a few cases double) Gaussian curves, using the method as described in Section \ref{sec:TSpecEmmission}. 

The wavelength calibration can be checked using the Na\,I doublet at 2206.2/2209.0 nm and the CO bandhead at 2295 nm. Emission lines may be wavelength shifted from radial and/or rotational gas kinematics.

\begin{figure*}
 \centering
 \includegraphics[width=1\linewidth]{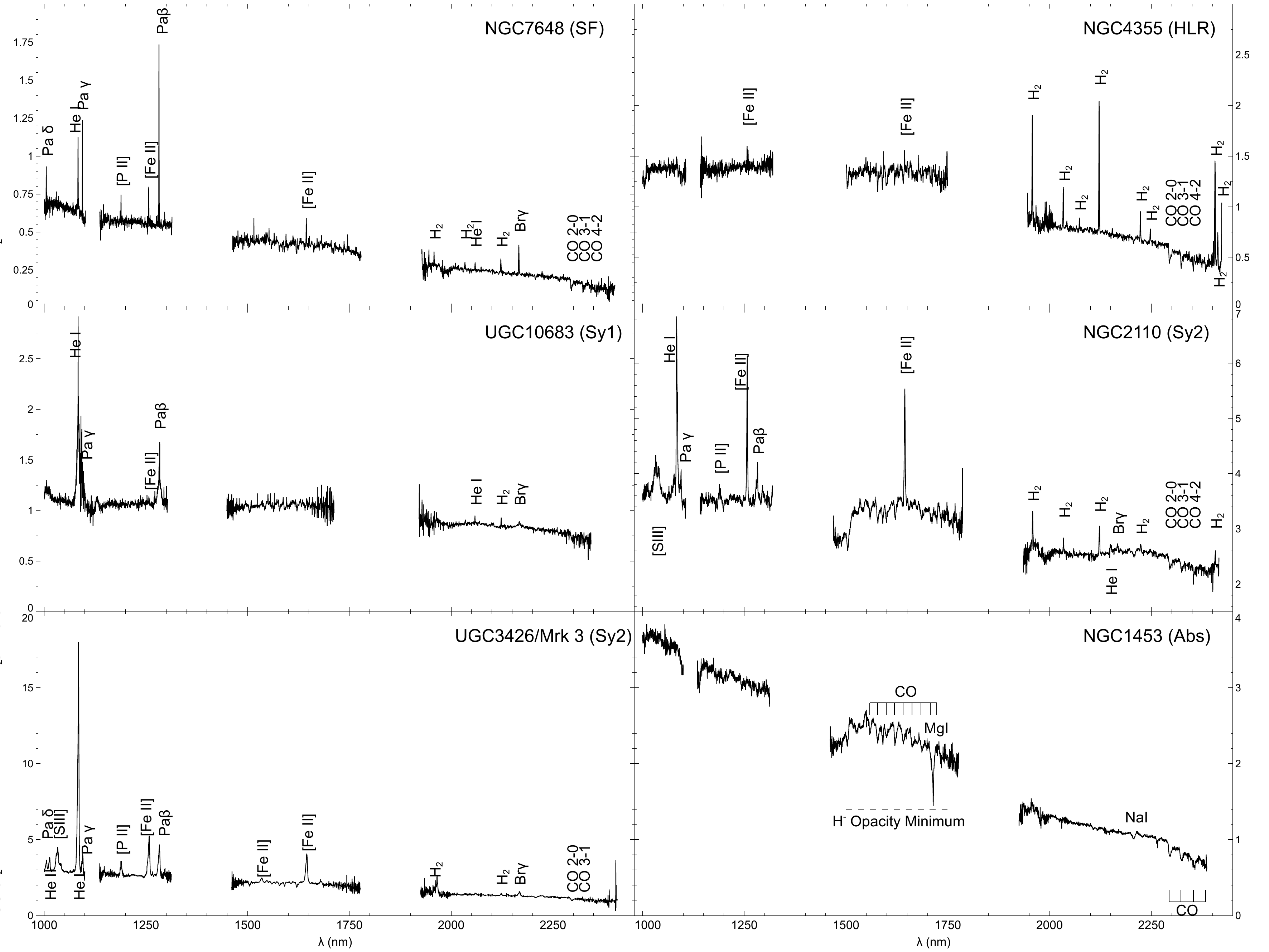}
 \caption{Example spectra, labelled with the galaxy name and activity type (SF - star forming, HLR - high line ratios, Sy1 - Seyfert 1, Sy2 - Seyfert 2, ABS - absorption lines only)}
 \label{fig:tspecdatapaper05}
\end{figure*}

The signal to noise, as given in Table \ref{tbl:TSpecAtlas02}, is derived using the \texttt{DER\_SNR} algorithm from the Space Telescope Science Institute (STScI)\footnote{\url{www.stecf.org/software/ASTROsoft/DER_SNR/}}. Example plots of the resultant spectra can be seen in \citetalias{Mould2012} (their Figures 1, 2 and 4) and in Figure \ref{fig:tspecdatapaper05}; this shows a variety of nuclear excitation types, star-forming (strong narrow hydrogen lines relative to other emission lines), Seyfert 1 (broad permitted lines of hydrogen and \HeI) and 2 (narrow permitted lines of similar strength to forbidden lines), HLR (hydrogen lines much weaker than \Fe{} and \Htwo), and purely absorption (i.e. no activity/emission lines). Figure \ref{fig:tspecdatapaper05} also labels the major emission lines; the purely absorption line sample (NGC~1453) labels the major absorption features; the molecular CO bandheads in the \textit{H} and \textit{K} bands, the ``bump'' in the range $\sim$1500-1750 nm from the minimum in the opacity of H$^{-}$ in cool stars \citep{Mason2015,Rayner2009} as well as many other absorption lines in the \textit{K}-band, including the prominent \MgI/CO line at 1711 nm. These absorption and emission lines are discussed in more detail in Section \ref{sec:TSpecAbsorption} and \ref{sec:TSpecEmmission}.
\subsection{Sample Characteristics}
\begin{figure*}
 \centering
 \includegraphics[width=1\linewidth]{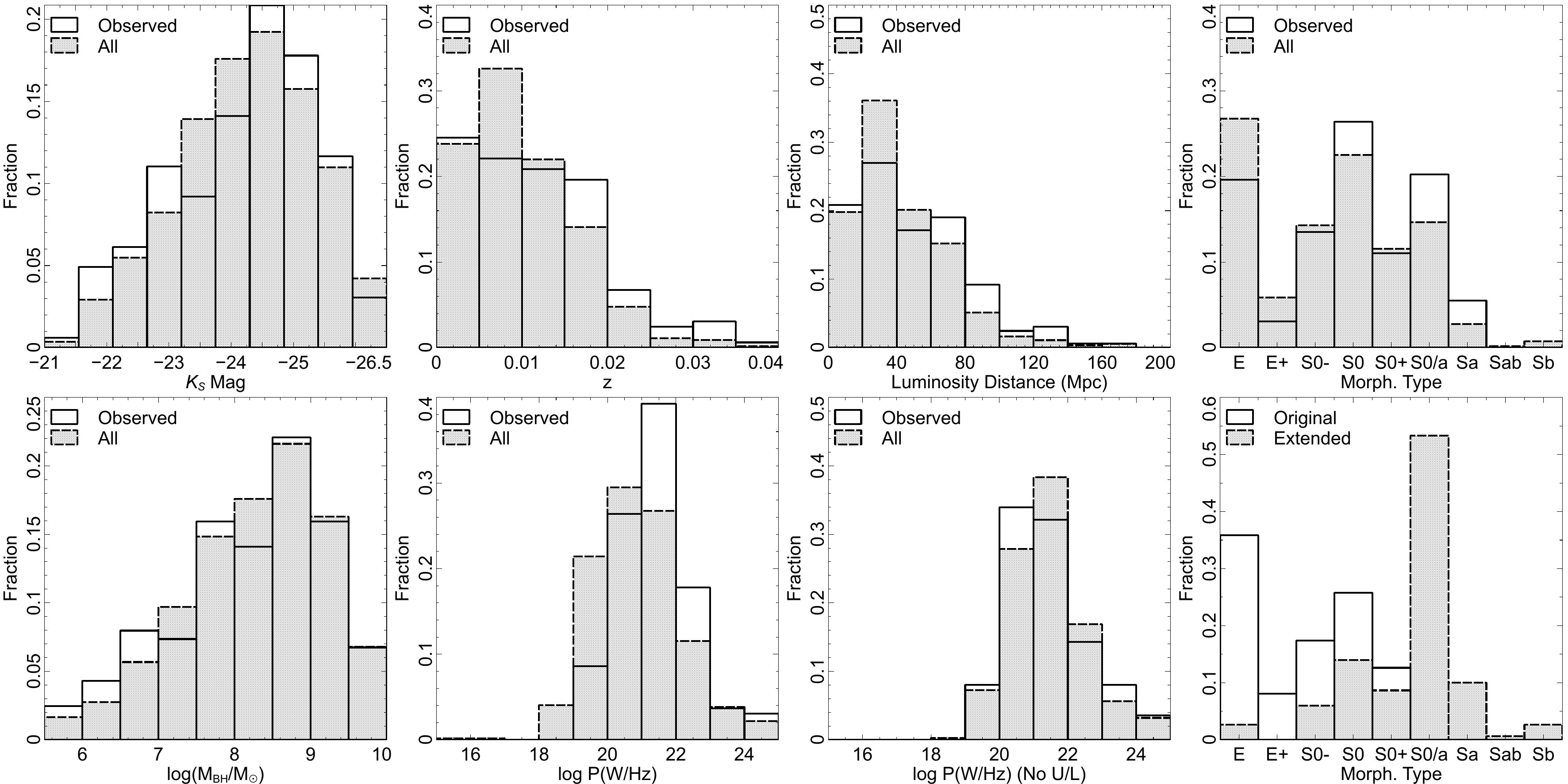}
 \caption{Histograms of absolute \textit{K}-band magnitude, redshift (z), morphological type and computed black hole mass, for the observed objects vs. the complete catalog.}
 \label{fig:tspecdatapaper09}
\end{figure*}

Figure \ref{fig:tspecdatapaper09} presents histograms of distributions of redshift (z), morphological type (both of observed vs. all objects, and original vs. extended/supplementary catalogs), \textit{K}-band absolute magnitude ($M_K$), luminosity distance, radio power (both including and excluding upper limits)  and estimated black-hole mass. The distributions of the observed objects are compared with the full catalog of 546 objects. In general, these match reasonably well, with a bias towards more objects of morphological type 0 (S0/a galaxies) observed than in the complete catalog. By selection, the original vs. extended/supplementary catalogs are ETGs (with no morphological types later than S0) vs. later type galaxies. We also observed objects that have a slightly greater $M_K$ luminosity (and thus computed black hole mass) and  radio power. When we exclude upper limits to the radio power, then the two sets are approximately the same. {For more details on how these parameters were obtained, see Section \ref{sec:TSpecRadioCatalog}}.

\section{Results and Discussion}
\subsection{Absorption Features}
\label{sec:TSpecAbsorption}
NIR spectral absorption features can complement optical studies of stellar populations of galaxies. The NIR presents some advantages over the optical regime \citep{Gasparri2021}; asymptotic giant branch (AGB) and red giant branch (RGB) spectral components can be isolated (the contribution of these populations outside of the \textit{K} band is negligible) and dust reddening is much reduced in highly obscured galaxies. Absorption line or band indexes can be used to determine stellar populations. These have been studied by e.g. \cite{Gasparri2021} (following \citealt{Cesetti2013} and \citealt{Morelli2020}). ETGs have specifically been studied by \cite{Cesetti2009} and \cite{Riffel2019a}.

\begin{figure*}
 \centering
 \includegraphics[width=0.8\linewidth]{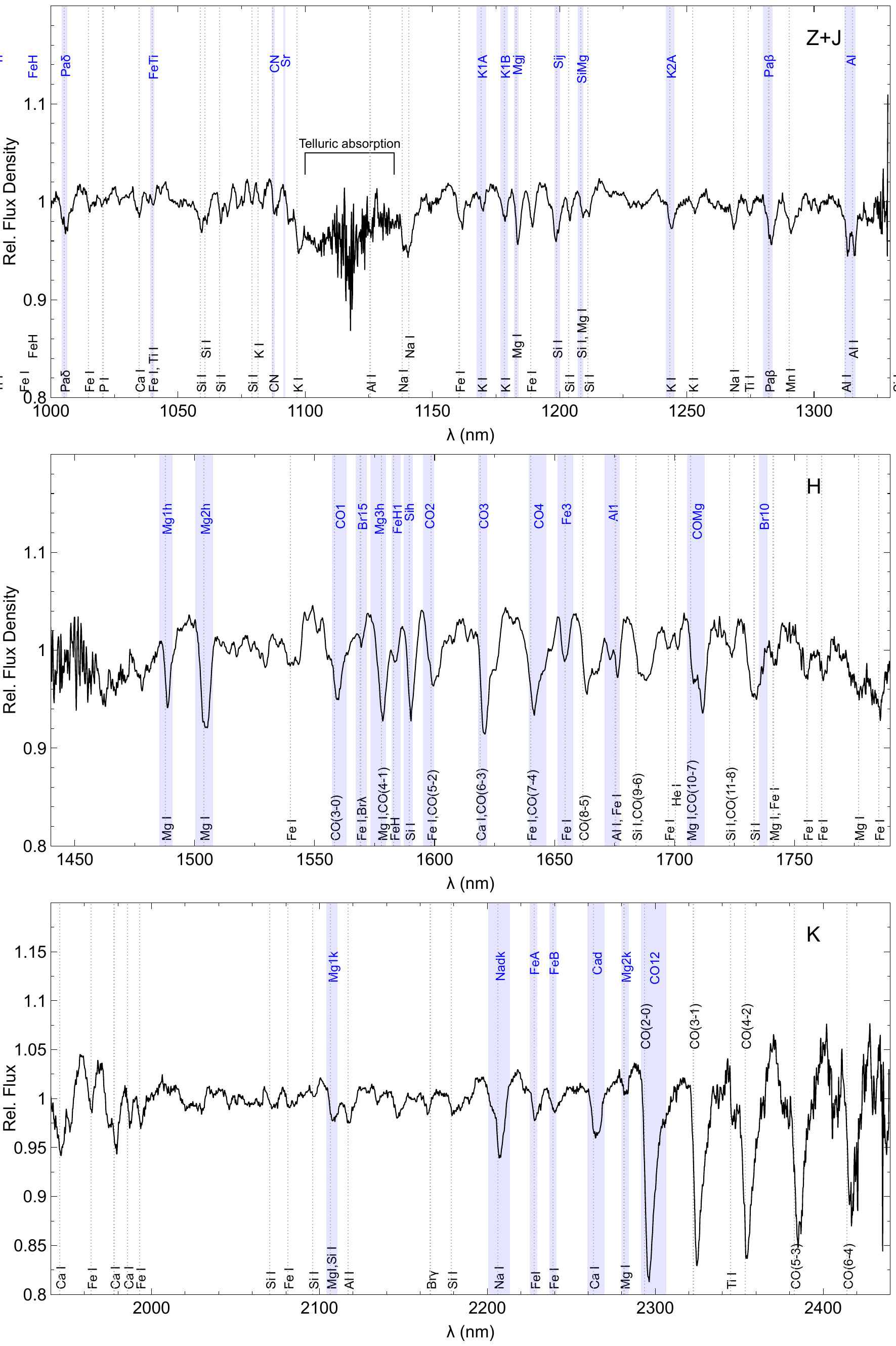}
 \caption{Stacked normalised spectrum (from objects with no emission lines) showing the main absorption features. The blue bands highlight the spectral indices, named with blue text, as discussed in the text.}
 \label{fig:tspecdatapaper01}
\end{figure*}

Studies of composite stellar systems (i.e. galaxies hosting an AGN) often need to subtract the contribution of the underlying galaxy \citep{Cesetti2009}. We use our data set to present an ``average'' spectrum of objects that do not show any emission lines (79/163 objects). We process the spectra as follows. Each spectrum was smoothed over a 100 pixel boxcar kernel, masking out the major emission-line regions (\HeI, \pag, \pab, \Fe, \Htwo{} and \brg), plus the CO bandheads above 2293 nm. This smoothed spectrum was divided through the original to create a continuum-normalised spectrum. We then corrected the dispersion solution to account for minor imperfections in the cataloged redshift, by measuring the wavelength of prominent absorption features in the \textit{K} and \textit{H} bands; we mainly used the \NaI{} feature at 2207.6 nm. We then averaged the spectra; the result is shown in Figure \ref{fig:tspecdatapaper01}. This plot has been broken up into the \textit{Z+J}, \textit{H} and \textit{K} bands. It shows many atomic absorption lines, including \FeI, \AlI, \CI, \CaI, \NaI, \MgI{} and \SiI. Identifications have been obtained from  \cite{Meyer1998}, \cite{Wallace2000}, \cite{Rayner2009}, \cite{Martins2013} and \cite{Gasparri2021}. The blue bands highlight those features used as spectral indices, from the work of \cite{Gasparri2021}. We note that the Paschen indices (\pab{} and \pag) are visible in this stacked spectrum, while the Brackett series is not present. The stacked spectrum also does not show any emission-line activity; in other words there is no low-level, residual emission in these objects.
It is well-known that the presence of an AGN dilutes the CO bandheads around 2300 nm, due to the rising hot dust continuum emission \citep{Riffel2009b,Riffel2017}. This corresponds to a spectral reddening (flattening), as measured by the \jk{} colour. We computed this colour from the average flux in a 10 nm window around the effective wavelength of the 2MASS \textit{J} (1295 nm) and \textit{K} (2159 nm) filters; this was converted to a magnitude using the NASA/IPAC magnitude/flux density converter\footnote{\url{https://irsa.ipac.caltech.edu/data/SPITZER/docs/dataanalysistools/tools/pet/magtojy/}} values of the zero-magnitude flux for each filter. We also calculated the CO12 line index using the methodology outlined in \cite{Cesetti2009}. 

\begin{figure}
 \centering
 \includegraphics[width=1\linewidth]{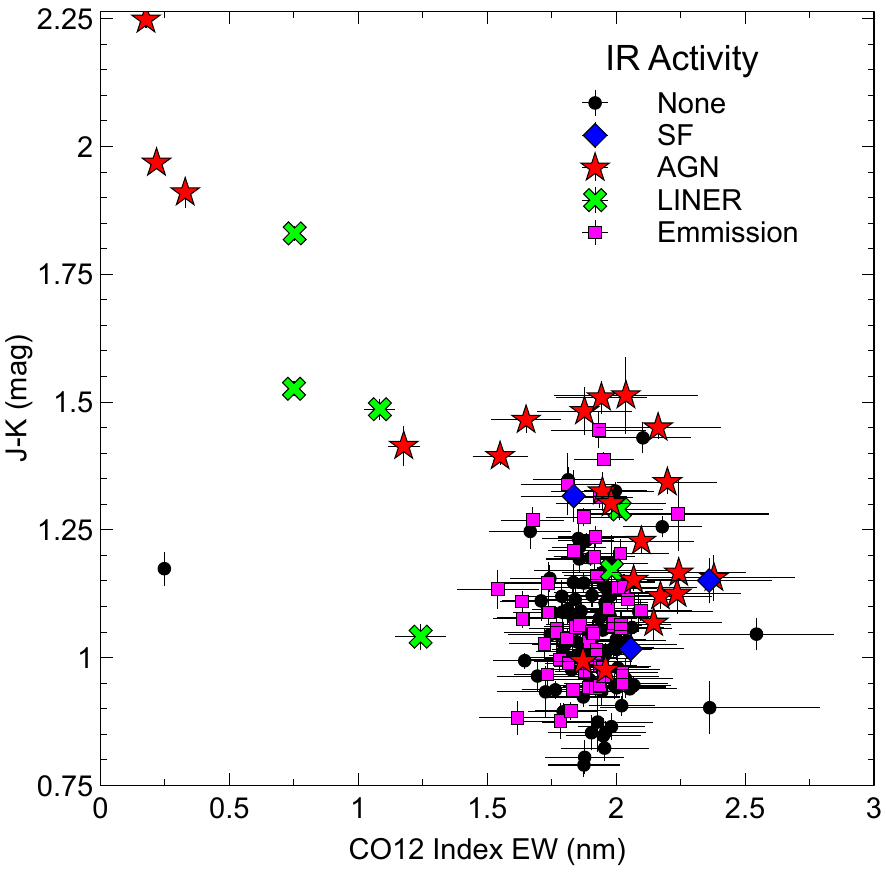}
 \caption{CO12 absorption line index vs. \jk{} colour, symbol and color coded by the IR nuclear activity. This shows a trend with nuclear activity producing a weaker CO12 index and a redder (flatter) continuum}
 \label{fig:tspecdatapaper11}
\end{figure}

Figure \ref{fig:tspecdatapaper11} plots the CO12 index equivalent width (in nm) against the \jk{} magnitude. The points are symbol and colour coded by the IR activity type (as derived in Section \ref{sec:TSpecNucActDiag} below). Most galaxies with either no activity or star formation are clumped together, but there is a visible trend with those diagnosed with AGN or LINER activities to have higher \jk{} values and/or a lower index. The 4 objects with \jk{} above 1.75 mag are Type 1 Seyferts (IC~450, NGC~5548, UGC~10683) or the Seyfert 2 NGC~2110 with powerful emission lines. The outlying point with a CO12 index = 0.25 and \jk{} = 1.2 is a spectrum with poor S/N. The ``Emission'' objects are those with any of \HeI, \Fe{} or \Htwo, without hydrogen lines to enable nuclear activity classification.

\subsection{Emission Features}
\label{sec:TSpecEmmission}
The spectra were inspected for emission lines and these were measured where found. The lines measured were hydrogen (\pab{} at 1282 nm and \brg{} at 2166 nm), forbidden iron \Fe{} at 1257 nm, \Htwo{} at 2121 nm and \HeI{} at 1083 nm.  Line identification was usually unproblematic, even with considerable noise. 

We used the \texttt{QFitsView}\footnote{\url{http://www.mpe.mpg.de/~ott/dpuser/qfitsview.html}} \citep{Ott2012} ``de-blending'' functionality to fit all emission lines; this allows for both Gaussian and Lorentzian function fits with single or multiple components. This requires some manual input from the user to set the initial estimates of continuum, height, centre and width and uses the GSL\footnote{\url{https://www.gnu.org/software/gsl/doc/html/index.html}} \textit{gsl\_multifit} routines, returning fit values and errors of each component (continuum slope, central wavelength, FWHM and flux). Table \ref{tbl:TSpecAtlas03} gives the flux measurement and errors for a sample of objects; the complete list is available online, as described below. All fluxes are given in units of 10\pwr{-16} \ecs.

The emission lines were usually fitted with a single Gaussian function. In some cases, the permitted hydrogen and \HeI{} lines were broad, indicating Seyfert 1 type activity (8/163 objects). In these cases, the lines were fitted with 2 or more Gaussian functions. In Section \ref{sec:TSpecEmmLineComps}, we analyse these broad lines as the combination of multiple components.

In the spectra of elliptical galaxies presented by \citetalias{Mould2012}, approximately 20\% of the complete \citetalias{Brown2011} catalog showed \brg{} emission, indicating star formation or nuclear activity. In this sample, extended to S0 objects, 50\% (82/163) show emission lines, 19\% (31/163) show hydrogen lines, with a further 15 having \HeI{} flux (without hydrogen emission), indicative of nuclear activity. 38 objects also showed just \Fe{} and/or \Htwo{} emission, indicating HLR activity. In general, hydrogen recombination emission lines are indicative of star formation and/or nuclear activity. In the SF case, the line emission comes from HII regions excited by ionizing photons from massive, young, hot stars, as well as from the stellar atmospheres; another source is from supernova remnants, where the line emission is mainly from the shocked gas. For the infrared lines \pab{} and \brg, single-stellar population models indicate, for an instantaneous starburst, that these have disappeared by 20 Myr; however, for continuous SF, these can persist for > 1 Gyr \citep[e.g. STARBURST99,][]{Leitherer1999}.

17 galaxies had activity classifications from \cite{Veron-Cetty2006} and from the SIMBAD Astronomical Database\footnote{\url{http://simbad.cds.unistra.fr/simbad/}}, but showed no measurable emission-line fluxes in the NIR. This can be attributed to poor signal to noise on the measured spectrum; on the other hand, the available optical spectra can be somewhat uncertain. 

\begin{table*}
\centering
\caption{Sample of emission-line fluxes for the main atomic and ionised species, in units of 10\pwr{-16} \ecs. \ebv{} is calculated from the \pab/\brg{} flux ratio, as discussed in section \ref{sec:TSpecExtinction}. (The full table is available online)}
\label{tbl:TSpecAtlas03}
\resizebox{\textwidth}{!}{%
\begin{tabular}{@{}lcccccc@{}}
\toprule
Object                  & He I             & \Fe             & \pab            & \Htwo{}        & \brg          & \ebv\\ 
&(1083 nm)&(1257 nm)&(1282 nm)&(2121 nm)&(2166 nm)&(mag)\\
\midrule
2MASX J20173144+7207257 & 34.5$\pm$3.3     & 23.3$\pm$2.2    & 41.8$\pm$1.9    & 10.7$\pm$1.2   & 12.7$\pm$1.5  & 1.16  \\
ESO483-12               & 20.4$\pm$6.9     & 24.0$\pm$9.9    & 36.3$\pm$6.0    & 17.1$\pm$1.4   & 13.1$\pm$1.3  & 1.93  \\
ESO507-25               & 56.7$\pm$32.6    & 25.4$\pm$10.0   & \dots           & 43.4$\pm$3.1   & \dots         &       \\
IC0051                  & 15.2$\pm$1.8     & 49.6$\pm$0.9    & 18.6$\pm$1.1    & 2.9$\pm$0.3    & 3.7$\pm$0.7   & 0.35  \\
IC0450                  & \dots            & 514.0$\pm$5.3   & 967.6$\pm$134.4 & 76.7$\pm$10.0  & 71.1$\pm$10.8 & -2.34 \\
IC0537                  & \dots            & 54.2$\pm$16.3   & \dots           & 52.8$\pm$5.2   & \dots         &       \\
IC0630                  & 4470.0$\pm$67.6  & 339.0$\pm$16.3  & 2010.0$\pm$37.1 & 37.4$\pm$2.9   & 536.0$\pm$9.6 & 1.13  \\
MCG-01-29-015           & \dots            & \dots           & \dots           & 21.6$\pm$1.0   & \dots         &       \\
MCG-02-33-017           & 23.6$\pm$4.2     & \dots           & \dots           & \dots          & \dots         &       \\
MCG+04-50-004           & 381.0$\pm$4.6    & 21.8$\pm$4.4    & 50.4$\pm$32.0   & 24.8$\pm$1.6   & \dots         &       \\
MRK612                  & 172.9$\pm$13.1   & 82.4$\pm$6.2    & 58.6$\pm$6.9    & 37.5$\pm$2.8   & 15.4$\pm$2.7  & 1.09  \\
NGC0051                 & \dots            & 28.7$\pm$6.7    & \dots           & 35.6$\pm$2.1   & \dots         &       \\
NGC0383                 & \dots            & 6.4$\pm$2.1     & \dots           & \dots          & \dots         &       \\
NGC1052                 & 389.0$\pm$24.9   & 342.0$\pm$16.0  & \dots           & 33.2$\pm$3.3   & \dots         &       \\
NGC1222                 & 753.0$\pm$6.3    & 81.7$\pm$3.5    & 493.0$\pm$7.0   & 18.7$\pm$2.7   & 167.0$\pm$5.2 & 1.76  \\
NGC2110                 & 814.0$\pm$38.8   & 588.0$\pm$12.3  & 59.0$\pm$7.1    & 107.0$\pm$35.6 & 23.3$\pm$3.9  & 1.89  \\
NGC4438                 & 114.0$\pm$28.5   & 215.0$\pm$18.3  & 37.5$\pm$5.2    & 174.0$\pm$7.4  & \dots         &       \\
NGC5273                 & 1125.0$\pm$51.2  & 31.0$\pm$3.4    & 7.5$\pm$4.9     & 32.2$\pm$1.8   & \dots         &       \\
NGC5548                 & 4810.0$\pm$76.6  & 22.8$\pm$5.7    & 58.1$\pm$7.5    & 11.4$\pm$1.9   & 10.5$\pm$2.5  & 0.13  \\
NGC7465                 & 707.0$\pm$81.0   & 213.0$\pm$10.9  & 178.0$\pm$12.0  & 84.7$\pm$7.2   & 23.0$\pm$5.3  & -0.80 \\
UGC3426                 & 4700.0$\pm$200.0 & 1210.0$\pm$36.0 & 809.0$\pm$36.1  & 42.8$\pm$4.5   & 174.0$\pm$8.7 & 0.45  \\
UGC5745                 & 41.9$\pm$9.1     & 62.9$\pm$5.8    & 51.2$\pm$6.0    & 16.0$\pm$3.1   & 15.7$\pm$2.3  & 1.49  \\ \bottomrule
\end{tabular}}
\end{table*}

\subsection{Multi-component Emission Lines}
\label{sec:TSpecEmmLineComps}
For those objects that show multiple components in their emission lines (either from the AGN BLR emission or from multiple NLR components with different velocities - 6/163 objects), we analyse the \HeI{} and \pab{} lines by fitting a double Gaussian curve. The results are given in Table \ref{tbl:TSpecAtlas07} below. Each object has different characteristics, with the \pab{} line in general being the most complex. In all cases the \HeI{} line could be fitted with single broad and narrow Gaussian curves. 

For NGC~5548 \citep[following][]{Schonell2017}, we fit the \pab{} with two broad and one narrow Gaussian curves (with the same for NGC~5273 and IC~450). The \pab{} velocity structures are not clear, with the fitted broad components not on the restframe wavelength. The narrow components are at restframe and a relative velocity of $\sim$ -2700 \kms{} (IC~450) and $\sim$ -2570 and +1400 \kms{} (NGC~5548); these can be interpreted as outflows. For IC~450, we were not able to fit the \HeI{} line, due to the imperfect subtraction of the telluric OH line at 1097.53 nm. NGC~3516 has 3 prominent \pab{} narrow lines superimposed on the  broad line. While UGC~3426 (Mrk~3) has broader lines than standard narrow-line width ($\sim$ 525 \kms), this is not a classic broad-line region, as the permitted and forbidden lines are the same width; the line profiles are caused by powerful outflows of ionised gas. In Table \ref{tbl:TSpecAtlas07}, where there are multiple components, the fluxes are the sum of all components.

\begin{figure}
 \centering
 \includegraphics[width=1\linewidth]{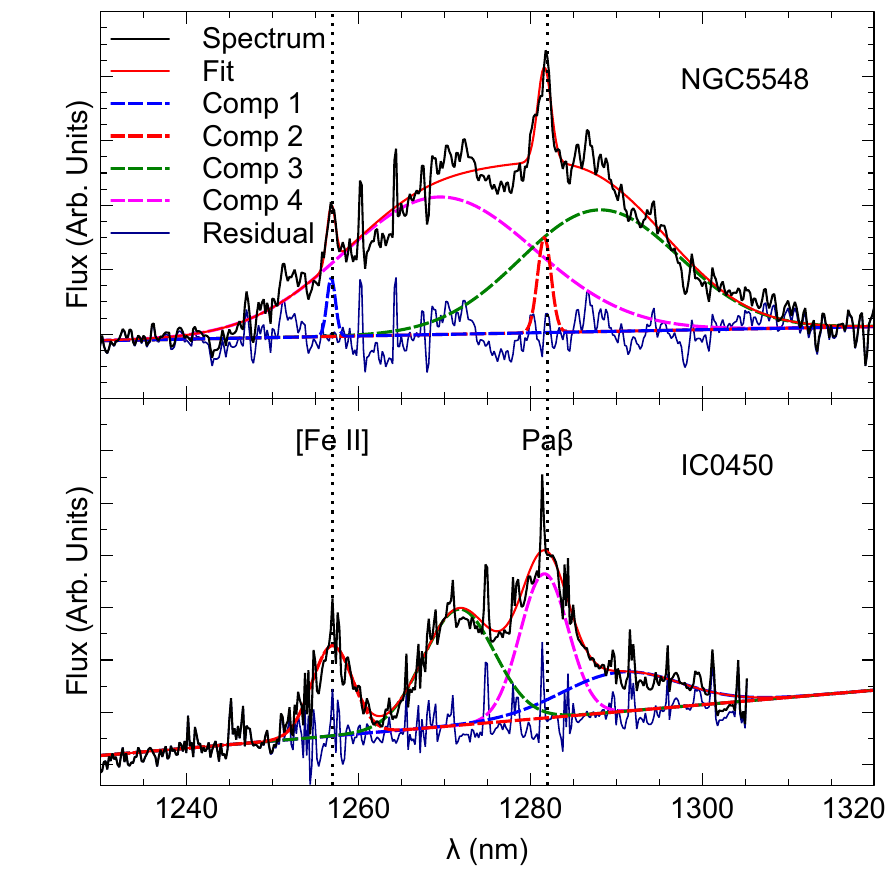}
 \caption{Examples of multiple Gaussian curve fits to Seyfert 1 line profiles for NGC~5548 and IC~450.}
 \label{fig:tspecdatapaper10}
\end{figure}

Figure \ref{fig:tspecdatapaper10} gives examples of Gaussian component fits to the broad \pab{} emission line for NGC~5548 and IC~450, with the object spectrum, the individual components and the residuals.
\begin{table*}
\centering
\caption{\HeI{} and \pab{} parameters for objects showing multiple components in emission. (This table is available online).}
\label{tbl:TSpecAtlas07}
\resizebox{\textwidth}{!}{%
\begin{footnotesize}
\begin{tabular}{lrccrccrccrccl}
\toprule
Object    & \multicolumn{3}{c}{\HeI~(b)}&\multicolumn{3}{c}{\HeI~(n)}&                  \multicolumn{3}{c}{\pab~(b)}  & \multicolumn{3}{c}{\pab~(n)}      & Notes \\
          & Flux$\star$         & $\lambda\dagger$ & W$\ddagger$& Flux            & $\lambda$ & W     & Flux             & $\lambda$ & W     & Flux            & $\lambda$ & W     &       \\ \midrule
NGC~3516  & 38.9$\pm$1.6 & 1083.5 & 14.3  & 6.8$\pm$0.5  & 1083.4 & 2.7   & 11.8$\pm$0.8 & 1280.5 & 13.1  & 1.3$\pm$0.2 & \dots  & \dots & A \\
NGC~5273  & 11.4$\pm$0.5 & 1084.3 & 17.2  & 1.0$\pm$0.1  & 1083.8 & 0.9   & 5.7$\pm$0.7  & \dots  & \dots & 0.4$\pm$0.1 & 1278.1 & 2.2   & B \\
NGC~5548  & 46.7$\pm$0.8 & 1084.3 & 27.8  & 7.2$\pm$0.2  & 1083.8 & 2.1   & 23.9$\pm$0.6 & \dots  & \dots & 0.6$\pm$0.1 & 1282.8 & 1.9   & C \\
NGC~7465  & 7.8$\pm$0.6  & 1083.4 & 4.6   & 2.8$\pm$0.2  & 1083.3 & 0.9   & 4.3$\pm$0.4  & 1281.6 & 6.0   & 1.8$\pm$1.1 & 1281.9 & 0.9   &   \\
UGC~10683 & 6.2$\pm$0.8  & 1083.7 & 11.7  & 1.8$\pm$0.2  & 1083.7 & 3.4   & 2.2$\pm$0.4  & 1283.2 & 9.8   & 0.4$\pm$0.1 & 1283.3 & 2.6   &   \\
IC~0450   & \dots        & \dots  & \dots & \dots        & \dots  & \dots & 18.0$\pm$3.0 & \dots  & \dots & 9.7$\pm$1.3 & 1281.6 & 6.6   & D \\ \bottomrule
\end{tabular}
\end{footnotesize}
}
\renewcommand{\arraystretch}{1.1}
\normalsize
\begin{flushleft}
\begin{tabular}{rl}
\textbf{Notes:}\\
$\star$&Flux in units of 10\pwr{-14}\ecs.\\
$\dagger$&Restframe wavelength of Gaussian fit in nm.\\
$\ddagger$&Full-width half maximum in nm.\\
A&\pab{} narrow line flux total of 3 components at 1279.03, 1282.10, 1287.18 nm.\\
B&\pab{} broad line 2 components at 1282.34, 1300.01 nm.\\
C&\pab{} broad line 2 components at 1270.37, 1289.09 nm.\\
D&\HeI{} line affected by telluric skyline. \pab{} broad line 2 components at 1271.69, 1290.52 nm.\\
\end{tabular}
\end{flushleft}
\end{table*}
\subsection{Nuclear Activity Diagnostics}
\label{sec:TSpecNucActDiag}
Nuclear activity for NIR emission line objects can be categorized by a diagnostic diagram \citep{Larkin1998,Rodriguez-Ardila2005}, where the log of the flux ratio \Htwo/\brg{} is plotted against that of \Fe (1257 nm)/\pab. This is analogous to the BPT diagrams \citep{Baldwin1981} commonly used in the optical regime \citep[e.g.][]{Kewley2006}. Following the updated limits from \cite{Riffel2013a}, the diagram is divided into three regimes for star formation/starburst (SF; dominated by H II regions), AGN (subjected to the radiation field from the accretion disk), and  high line ratio (HLR) excitation, where shocks from supernovae (SNe) and evolved stellar outflows dominate. We follow \cite{Riffel2021a} for the HLR nomenclature, as it encompasses transition objects, LINERs and supernovae remnants.  It is noted that nuclear activity can be apparent even though the galaxy may not be classified with a catalogued activity type from \cite{Veron-Cetty2006}. 

\begin{figure}
 \centering
 \includegraphics[width=1\linewidth]{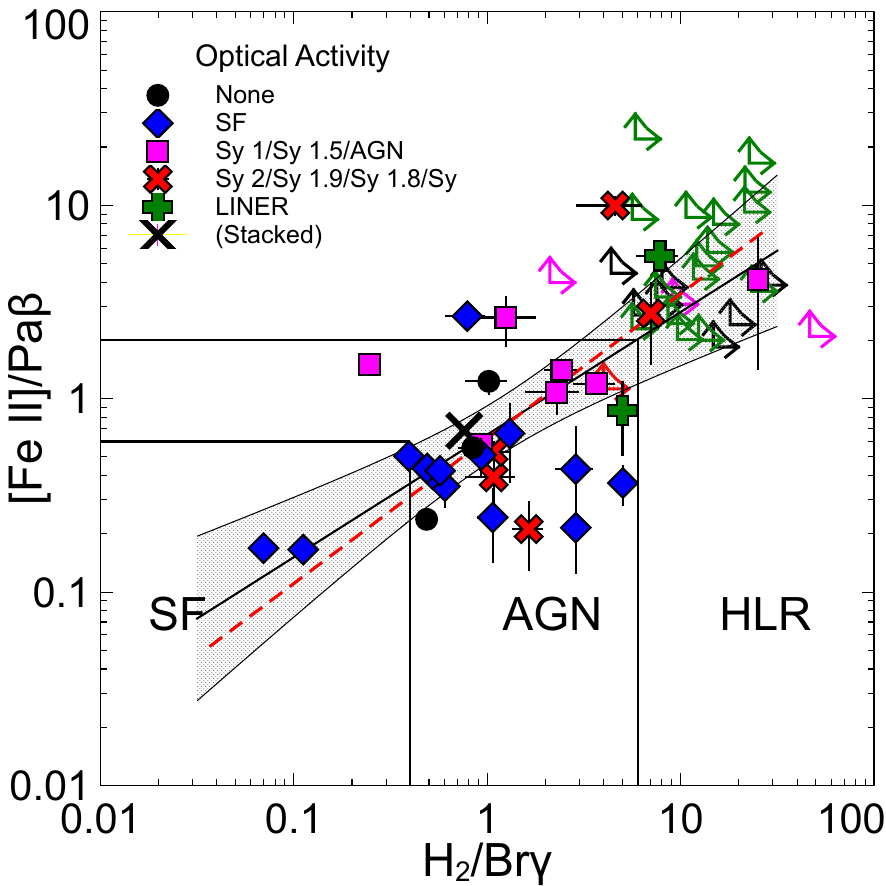}
 \caption{Infrared excitation diagram with the regimes from \cite{Riffel2013a} delineated, as discussed in the text. The emission line flux ratios \Fe/\pab{} vs \Htwo/\brg{} are plotted for those objects with measurable flux for all 4 lines. The optical activity type is coded by a symbol and colour as given in the legend, and matches the IR diagnostic well. The diagnostic point from stacking all emission-line spectra is the black cross symbol (Stacked). 10 galaxies optically classified as SF are revealed as ``hidden'' AGN or LINERs in the IR diagnostic. The correlation is similar to that found by \cite{Riffel2013a}; their fit (dashed red line) compares to that found here (orthogonal linear fit as a black line with 95\% confidence interval). Objects with no H lines are marked with upper limit arrows, using the same colour code for optical activity type; these are mostly LINERs.}
 \label{fig:tspecdatapaper03}
\end{figure}

These ratios are plotted in Figure \ref{fig:tspecdatapaper03} for those objects with measurable flux for all 4 lines. 26 of the 163 objects in our sample are detected with all these emission lines. A further 5 galaxies showed no \brg{} emission, but it could be calculated from the \pab{} flux using the \cite{Hummer1987} canonical ratio of \brg/\pab~=~5.85 (for $T_e = 10^4$K and $n_e = 10^{3}$ cm\pwr{-3}); this assumes no significant extinction. For those galaxies with multiple component broad-line emission, we used the narrow \pab{} and \brg{} components. Three of these galaxies (NGC~3516, NGC~5273 and NGC~5548) showed no broad \brg{} emission with weak or no narrow emission; given the \pab{} fluxes, these should have been visible. Over the standard range of ISM temperatures (500-30000 K) and densities ($n_e = 10^{2}-10^{7}$ cm\pwr{-3}), the \pab/\brg{} ratio varies $\sim$5-6.5. Higher-level hydrogen line emission can be suppressed in the extreme conditions of the AGN \citep{McAlary1986,Hubbard1985}, e.g. at $T_e = 10^3$K and $n_e = 10^{14}$ cm\pwr{-3}, the ratio is $\sim$33.5 \cite[e.g.][for Mrk~231]{Cutri1984}. For UGC~3426 (Mrk~3), we use the full flux value, as the multiple components and broad lines are caused by powerful outflows (as seen in the similar line widths for narrow lines e.g. \Fe{} and \Htwo).

Of these 31 galaxies, 4 are classified in the IR regimes as SF, 24 as AGN and 3 as HLR. The corresponding optical classification is plotted for each object as symbols and colours. The general correlation between the log(\Htwo/\brg) and log(\Fe/\pab) is apparent; the orthogonal linear fit is:
\begin{equation}
    \log([Fe II]/Pa\beta) = 0.634 \log(H_{2}/Br\gamma) - 0.186
\end{equation}
This is plotted in Figure \ref{fig:tspecdatapaper03} with 95\% confidence bands. This is comparable with the fit found by \cite{Riffel2013a} (their Figure 8, plotted on Figure \ref{fig:tspecdatapaper03} as a dashed red line). 4 objects have no optical classification, 13 are SB/HII/SBG, 5 are Sy1/Sy1.5, 6 are Sy1.9/Sy2 and 3 are LINERs. 

The plot also includes those objects with no visible hydrogen lines. These are plotted using the upper limits of \pab{} and \brg. These are mostly in the HLR region and are colour-coded with optical activity type. The great majority of these are LINERs.

Originally, 11 of these galaxies had no classification from \cite{Veron-Cetty2006} or NED (the NASA/IPAC Extra-galactic Database)\footnote{\url{http://ned.ipac.caltech.edu/}}. 7 of these were found to have optical spectra available from NED; the diagnostic lines (\Ha, \Hb, \NII, \SII{} and \OIII) were measured and the object classified according to the \cite{Kewley2006} diagrams, with 5 SF/HII and 2 Sy galaxies. 6 galaxies are classified by SIMBAD as "EmG", i.e. emission-line galaxies. These were also re-measured and found to be 4 SF, 1 Sy and 1 LINER.

In general, the optical and IR diagnostics are reasonable well matched, however, it is not perfect. 10 (of 14) optical SF/HII galaxies are in the IR AGN or HLR regimes; this is most likely due to the dust obscuration of the nucleus which is penetrated in the IR revealing the AGN; the optical SF diagnostic would then be from unobscured SF regions away from the nucleus. One galaxy (NGC~7465) was optically classified as a LINER, but shows broad \HeI{}, \pab{} and \brg{} emission lines. Measurement of its NED spectra re-classifies it as a Seyfert 2, which is then revealed as Seyfert 1 in the IR. The spectra of all the optical SF galaxies were checked as before, and found to be truly SF, except for NGC~3593, which was reclassified as a Seyfert 2. The changed classifications are flagged in the complete catalog, as described in Section \ref{sec:TSpecRadioCatalog}.

Similarly, two galaxies (NGC~5273 and NGC~7465) change classifications from Seyfert 1.9 and Seyfert 2 (respectively) to full-blown Seyfert 1 galaxies. Similar examples are found in \cite{Reunanen2002}.

We also plot the overall excitation diagnostic for all 31 galaxies, by stacking the spectra as described in Section \ref{sec:TSpecAbsorption}. This plotted as a black cross (Stacked), located in the AGN regime.

Objects with no H lines are marked with upper limit arrows, using the same colour code for optical activity type. The ratios are calculated using the upper limits of \pab{} and \brg, derived as described before.
\subsection{Molecular Hydrogen}
\label{sec:TSpecMolHydrogen}
Molecular hydrogen (\Htwo) is very important in the star formation context of AGN activity, since it is the basic building block for stars. In the \textit{K}-band, there are a series of rotational-vibrational emission lines, which can be used to examine the excitation mechanism for \Htwo; this can be either UV photons (fluorescence) from star formation and/or AGN continuum emission \citep{Black1987}, shocks from SNe, AGN outflows, star formation winds \citep{Hollenbach1989} or X-rays from the AGN irradiating and heating dense gas \citep{Maloney1996}. The distribution and excitation mechanisms for individual galaxies has been studied extensively \citep[e.g.][]{Mouri1994,Quillen1999,Davies2003,Rodriguez-Ardila2005,Wilman2005,Riffel2008,Riffel2021b}, especially in the context of active galaxies. Here we examine the bulk properties of the molecular gas from our sample.

Following the method outlined in \cite{Wilman2005}, for gas with density $n_T > 10^{5}$~cm\pwr{-3}, the thermal (collisional) temperatures can be estimated. The occupation numbers of the excited ro-vibrational levels of the \Htwo{} molecule will be in thermal equilibrium at a temperature $T_{exc}${} equal to the kinetic temperature of the gas. This leads to the relationship:
\begin{equation}
\ln{N_{upper}} = \ln\left( \dfrac{F_{i}~\lambda_{i}}{A_{i}~g_{i}}\right)  = constant - \frac{T_{i}}{T_{exc}}
\end{equation}
where $\ln(N_{upper})$ is the occupation number of the upper level of the \Htwo{} transition, $F_i$ is the flux of the \textit{i}th \Htwo{} line, $\lambda_i$ is its wavelength, $A_i$ is the spontaneous emission coefficient, $g_i$ is the statistical weight of the upper level of the transition and $T_i$ is the energy of the level expressed as a temperature. This relation is valid for thermal excitation, under the assumption of an \textit{ortho:para} abundance ratio of 3:1. The values of $\lambda_i$, $A_i$, $g_i$ and $T_i$ were obtained from the Gemini observatory online data resource ``Important H2 Lines''\footnote{\url{https://www.gemini.edu/observing/resources/near-ir-resources/spectroscopy/important-h2-lines}}, with the values updated from \cite{Roueff2022}\footnote{\url{http://cdsarc. u-strasbg.fr/viz-bin/cat/J/A+A/630/A58}}. 

\begin{figure*}
 \centering
 \includegraphics[width=0.8\linewidth]{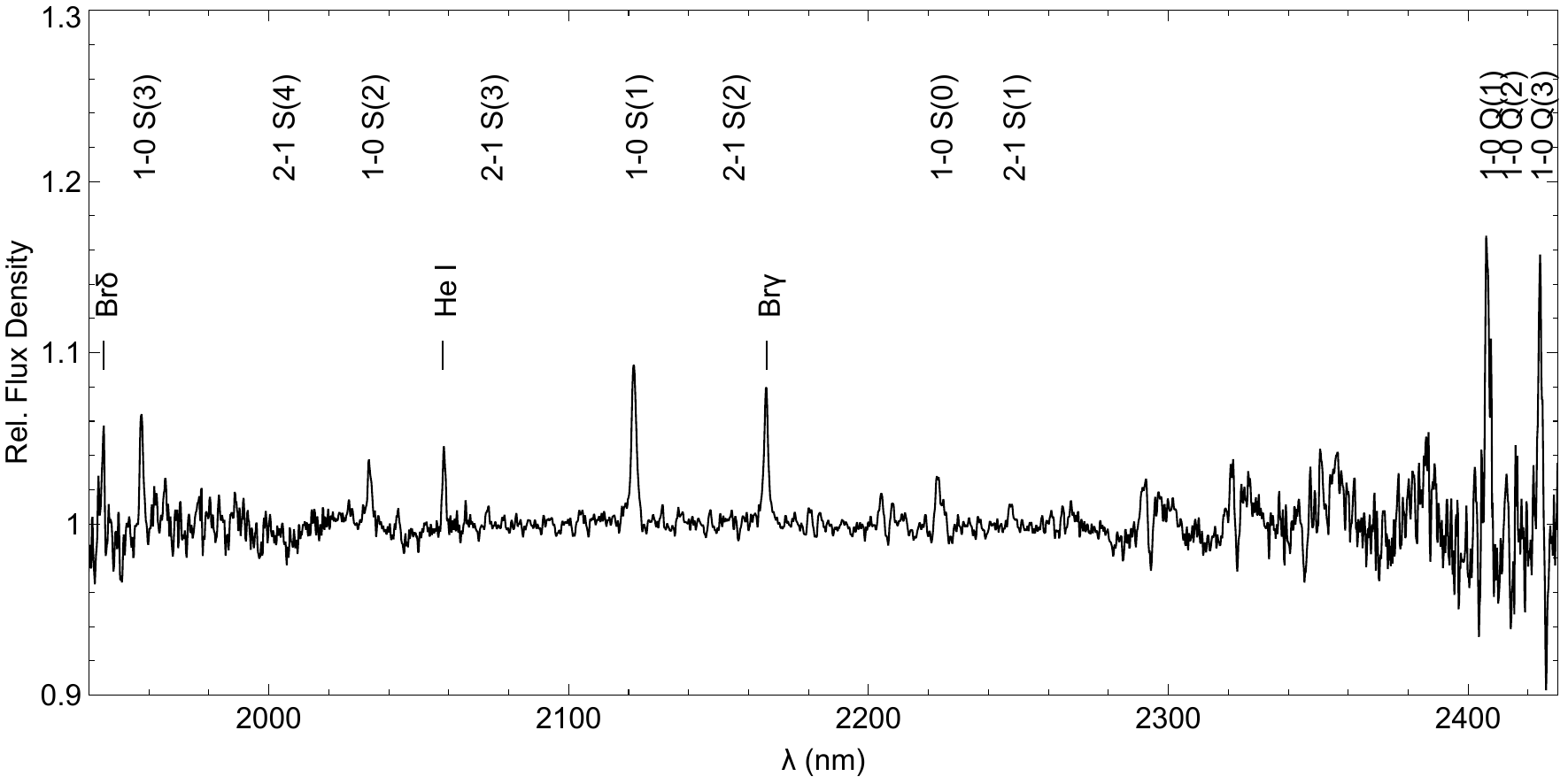}
 \caption{Stacked normalised spectra that have non-zero \Htwo{} flux, weghted by the continuum 2130-2145 nm. The emission lines are marked with the transition notation, including the prominent \brg, \brd{} and \HeI{} lines. The 1-0 S and Q lines are prominent, with the higher excitation 2-1 S lines much weaker.}
 \label{fig:tspecdatapaper04}
\end{figure*}

69 galaxies (42\% of 163) show \Htwo{} 2121 nm emission; to find an ``average'' temperature we stacked all the spectra, each normalised as described in section \ref{sec:TSpecAbsorption}. Each spectrum was weighted in the stacking process by the continuum flux, averaged over the wavelength range 2130-2145 nm. This is then divided though by the stacked absorption spectrum (see Section \ref{sec:TSpecAbsorption}) to remove absorption-line features. The resulting \textit{K}-band emission line sequence is shown in Figure \ref{fig:tspecdatapaper04}. 

The lines are marked with their transition notations; for a transition $\nu_{u},J_{u} \rightarrow \nu_{l},J_{l}$ (where $\nu_{u},J_{u};\nu_{l},J_{l}$ are the upper (\textit{u}) and lower (\textit{l}) level of the vibrational and rotational quantum numbers), is notated $\nu_{u}-\nu_{l}$ X($J_{l}$), where ``X'' is one of S, Q or O, depending whether $J$ is changed by -2, 0 or +2, respectively. For example, for the \Htwo{} 2121.8 nm line, the transition is 1-0 S(1), with $\nu_{1},J_{3} \rightarrow \nu_{0},J_{1}$.

In spectrum plot (Figure \ref{fig:tspecdatapaper04}), the hydrogen  \brg~(2166 nm), \brd~(1945 nm) and \HeI~(2058 nm) emission lines are also visible.

\begin{figure*}
 \centering
 \includegraphics[width=1\linewidth]{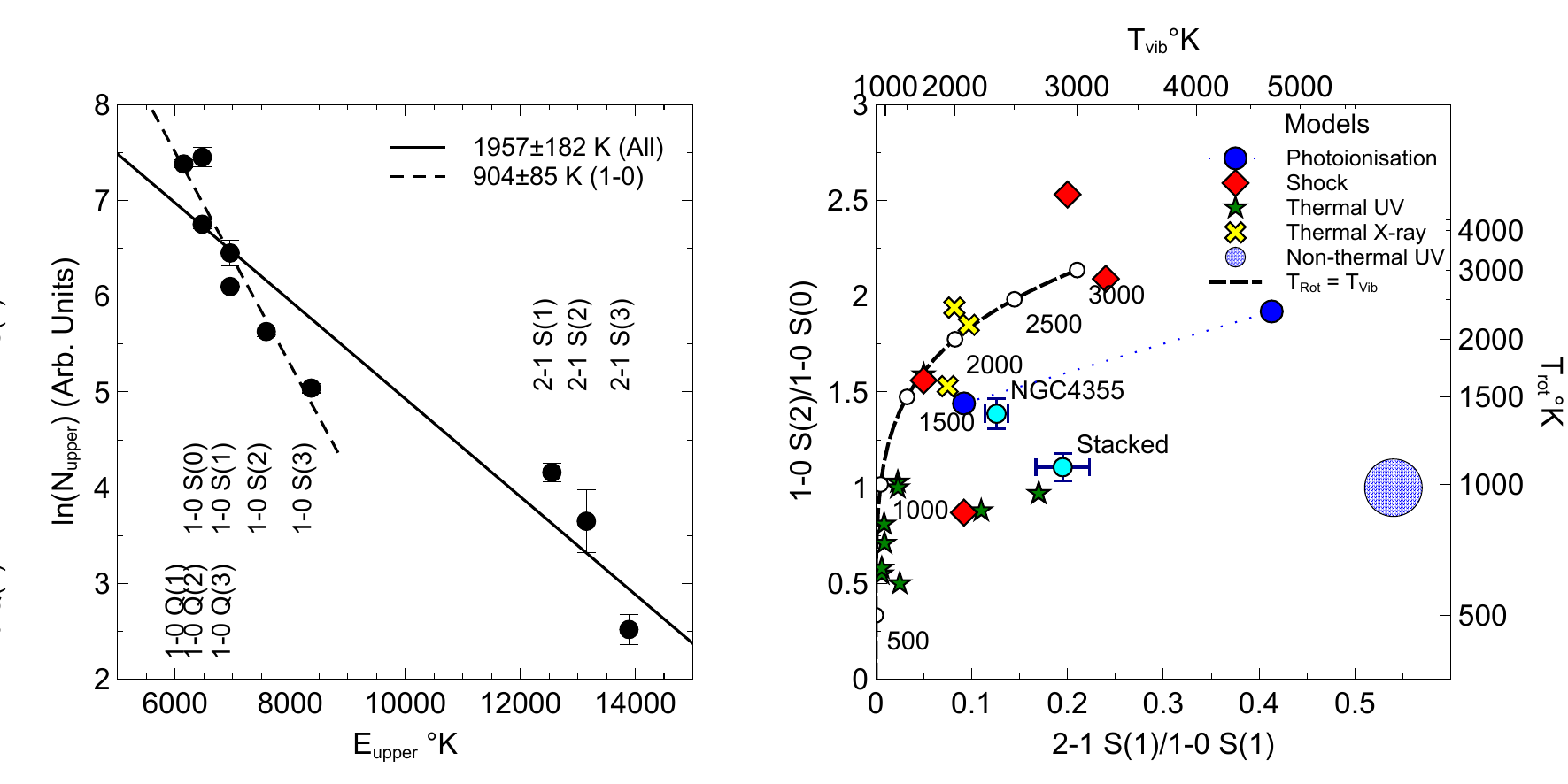}
 \caption{Left panel: \Htwo{} temperature plot. The value of the inverse slope of the relationship between ln(N$_{\rm{upper}}$) and E$_{\rm{upper}}$ is the excitation temperature (T$_{\rm{exc}}$) of \Htwo. We distinguish between all 2-1 and 1-0 transitions, vs. only those of the 1-0 transition. The transition notations are labelled are labelled, and the slopes with their corresponding temperatures given in the legend. Right panel: Excitation mechanism plot after \citep{Mouri1994} and \cite{Riffel2021b}, showing that the excitation is a mixture of thermal UV and shock mechanisms. Various excitation models are plotted, as described in the text.}   
 \label{fig:tspecdatapaper02}
\end{figure*}

Measuring the relative fluxes from the $\nu$=1-0 and 2-1 sequences, we fit a straight line to the calculated $\ln(N_{upper})$ plot; the negative inverse of the slope gives the temperature.   Fitting the values (as shown in Figure \ref{fig:tspecdatapaper02}), gives a temperature of $1957\pm182$ K. If we just fit the low-excitation lines ($\nu$=1-0), the temperature is lower ($904\pm85$ K). This can be due to several factors; (1) the high excitation lines ($\nu$=2-1) are weak, with corresponding uncertainties, (2) the warm \Htwo{} gas is not homogeneous with a smaller gas mass being hotter and (3) \cite{Davies2003,Davies2005} showed that the low-excitation levels may be thermalized, but the higher levels can be overpopulated owing to fluorescent excitation by far-ultraviolet photons.

We can examine whether various parameters affect the \Htwo{} temperature, by splitting the spectra into 2 bins. The results are given in Table \ref{tbl:TSpecAtlas09}. As the measurement of the 2-1 transitions is uncertain due to weak emission lines, we will concentrate on the temperature computed from the 1-0 transitions only. For radio power, the bin break is at log P (W Hz\pwr{-1}) = 21.5. Contrary to expectations, the higher radio power group has a significantly lower temperature than the lower power group; for both all and 1-0 transitions measurements, the 2-sided t-test for the slopes gives a significance $p = 0.0006$ and $p < 0.0001$ respectively. There is no difference in temperature when we split by \Htwo{} luminosity with the break at log L \es = 38.5 ($p = 0.29$ and $p = 1$). Table \ref{tbl:TSpecAtlas09} also gives the result for the galaxy with the highest \Htwo{} flux, NGC~4355, which has a somewhat higher temperature than the averaged value.

These temperatures are lower than those found for Seyfert galaxies observed with integral field spectroscopy (IFS), in the range 2100-2700 K (e.g. \citealt{Riffel2015,Storchi-Bergmann2009,Riffel2014a,Riffel2011,Riffel2010a,Durre2018a}). This is probably due to the spectroscopic slit encompassing cooler molecular regions than the higher resolution IFS observations, plus a large fraction of the galaxies not being AGNs.

\begin{table*}
\centering
\caption{\Htwo{} temperatures, measured from \textit{K}-band emission line sequence. As well as for all galaxies with \Htwo{} emission (weighted by the continuum flux), the objects are binned by low/high radio power and low/high \Htwo{} luminosity. NGC~4355 is included as the brightest \Htwo{} flux object. The table shows the object count, the temperatures for both the 1-0 and 2-1 transitions and the 1-0 transitions only, and the T$_{Vib}$ and T$_{Rot}$ temperatures (all in K).}
\label{tbl:TSpecAtlas09}
\normalsize
\begin{tabular}{@{}lccccc@{}}
\toprule
Method               & \# & Temp. (All)   & Temp. (1-0)  & T$_{Vib}$       & T$_{Rot}$      \\ \midrule
All& 69    & 1957$\pm$182 & 904$\pm$85  & 2882$\pm$408  & 996$\pm$69   \\
Radio power log P < 21.5    & 33    & 1950$\pm$144 & 882$\pm$65  & 2710$\pm$568  & 1028$\pm$206 \\
Radio power log P >= 21.5   & 36    & 1690$\pm$124 & 642$\pm$19  & 2097$\pm$809  & 710$\pm$91   \\
\Htwo{} luminosity log L < 38.5  & 34    & 1985$\pm$203 & 737$\pm$66  & 3324$\pm$1075 & 775$\pm$178  \\
\Htwo{} luminosity log L >= 38.5 & 35    & 1855$\pm$270 & 773$\pm$50  & 2012$\pm$379  & 970$\pm$101  \\
NGC4355              & 1     & 2080$\pm$91  & 1369$\pm$68 & 2357$\pm$223  & 1384$\pm$78  \\ \bottomrule
\end{tabular}
\end{table*}
The dominating excitation mechanism of \Htwo{} can be estimated and the contributing fractions of different mechanisms can be constrained \citep{Busch2016}. We use the methods of \cite{Mouri1994}, calculating the rotational and vibrational temperatures from the \mbox{[1--0~S(0)]/[1--0~S(2)]} and \mbox{[2--1~S(1)]/[1--0~S(1)]} flux ratios, respectively. Figure \ref{fig:tspecdatapaper02} right panel (after \citealp{Mouri1994} and \citealp{Krabbe2000}), plots excitation models, following \cite{Riffel2021b} (models from \citealp{Kwan1977,Black1987,Sternberg1989,Draine1990,Mouri1994,Smith1995,Davies2003,DorsJr2012}, see \citealt{Riffel2021b} Figure 8 for details). The ratios from the flux weighted measurements of all objects with  \Htwo{} emission is plotted, along with that for NGC~4355, which has the highest \Htwo{} flux of the sample. These are both in the thermal regime, with a mixture of UV photons and shocks as the driving mechanisms. 
\subsection{Extinction}
\label{sec:TSpecExtinction}
We can use the emission-line ratio \pab/\brg{} as an extinction diagnostic. In general, the extinction \kms{} is calculated from the ratio of the flux of two hydrogen species using the formula:
\begin{equation}
\label{eqn:TSpecextnc}
    E_{B-V}~=~\alpha~\log\left(\frac{\beta}{f_{1}/f_{2}}\right)
\end{equation}
where $f_{1}, f_{2}$ are the fluxes of species 1 and 2 (in this case \pab{} and \brg), $\alpha$ is a constant depending on the extinction law and the wavelengths of the two species and $\beta$ is the intrinsic ratio of the fluxes given the physical conditions of the emitting region. We use the canonical ratio (as above) of $\beta = 5.86$ and the \cite{Cardelli1989}  extinction law, giving $\alpha = 6.07$. The results are given in Table \ref{tbl:TSpecAtlas03}, for 25 objects where both \pab{} and \brg{} were measurable. The values were corrected for galactic extinction, using data from the IRSA Galactic Dust Reddening and Extinction service\footnote{\url{https://irsa.ipac.caltech.edu/applications/DUST/}} using the \cite{Schlafly2011} values; these were in the range 0.014-0.36 mag \ebv. The extinction varies from 0.13-2.58 mag \ebv; however 3 objects display anomalous ``negative'' extinctions. These are Seyfert 1 objects (IC~450, NGC~7465 and UGC~10683) where \brg{} is suppressed due to the extreme gas densities found in the BLR; this is further discussed in Sections \ref{sec:TSpecEmmLineComps} and \ref{sec:TSpecNucActDiag}. 
\subsection{NIR and Radio Emission Correlations}
\begin{figure*}
 \centering
 \includegraphics[width=.8\linewidth]{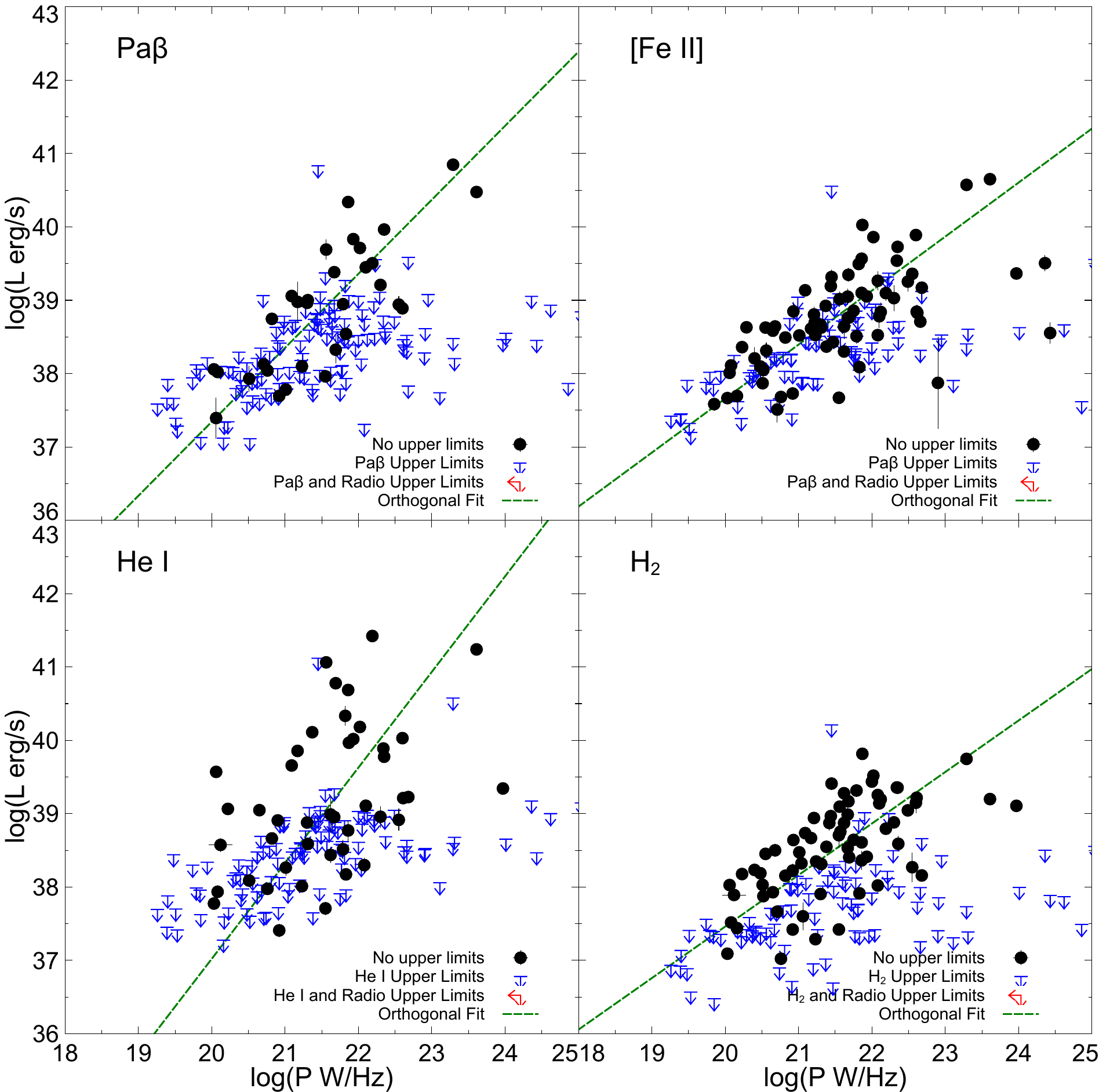}
 \caption{Correlation of radio power (X-axis) with NIR emission-line luminosities (Y-axis). The particular species is given in each panel, with black dots denoting firm measurements, blue symbols represent upper limit on the NIR fluxes and red symbols where both radio and NIR emission has upper limits. The orthogonal linear regression fit is also plotted.}
 \label{fig:tspecdatapaper06}
\end{figure*}

We explore the correlations between the radio and various emission-line luminosities. Figure \ref{fig:tspecdatapaper06} shows the plots of the radio luminosity on the X-axis vs. each emission line luminosity (Y-axis) as labelled on each panel (black points). We also include the plot for the emission-line upper limits (blue upper-limit arrow markers); these limits are calculated from the 3 $\sigma$ standard deviation over a 5 nm range around the line wavelength, with the continuum fitted locally with a third-order polynomial. In a few cases, both emission line luminosity and radio power have upper limits; these are marked with red symbols. Note there are no cases where there is a measured emission-line luminosity with a radio power upper limit. We plot the orthogonal linear regression fit. We note that while the \pab{} and \HeI{} fits have slopes $\ge$ 1, the \Fe{} and \Htwo{} fits are sub-linear. For molecular hydrogen, specifically, this means that at higher radio power, the AGN is emitting relatively less thermal energy from UV and X-rays and more radio power. This could provide an explanation of the lower \Htwo{} temperatures found at higher radio power, as noted in Section \ref{sec:TSpecMolHydrogen}.

While there is a trend with radio power of the emission-line luminosities, the upper limits are also, in general, showing the same trend; we therefore cannot couple them directly. While radio emission is always present (by sample selection), NIR line emission is not. This indicates that the duty cycles of SF and AGN activity are not synchronised. This has been explored in simulations, showing AGN activity triggering SF  \citep{Zubovas2013,Zubovas2019b}, as well as in the reverse direction \citep[e.g.][]{Schartmann2017}, where Toomre-instability induced nuclear SF can cause AGN activity though stellar outflows. \cite{Davies2007a,Davies2009} observe moderately recent (10–300 Myr) starbursts their sample of AGNs, deducing episodic periods of SF and positing a 50–100 Myr delay between SF and AGN activity onsets. They concluded that OB stars and supernovae (SNe) produce winds that have too high velocities to feed the AGN, but that evolved AGB stellar winds with slow velocities can be accreted efficiently onto the super-massive black hole (SMBH).

\cite{Ogle2010} found far infrared (FIR) warm molecular hydrogen emission from 17/55 (31\%) of a sample of radio galaxies, which compares to 42\% (69/163) of our sample. They suggest that their class of radio-selected molecular hydrogen emission galaxies (radio MOHEGS) is powered by radio-jet feedback in the form of kinetic energy dissipation by shocks or cosmic rays. However, their galaxies are nearly all members of clusters or close pairs (contrary to our sample) suggesting the \Htwo{} is delivered in galaxy collisions or cooling flow.

\begin{figure*}
 \centering
 \includegraphics[width=.8\linewidth]{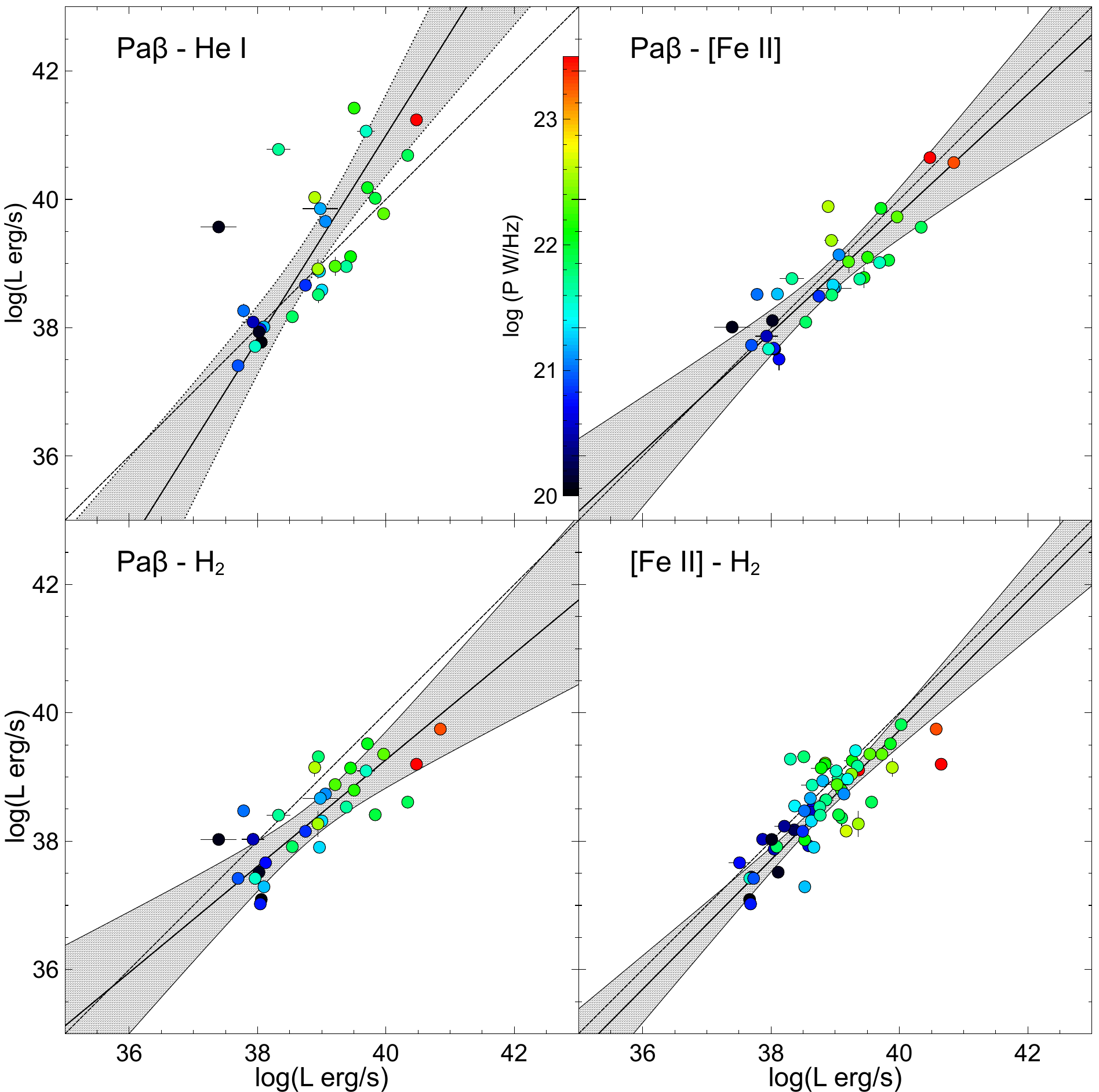}
 \caption{Correlations between NIR species luminosities. The X- and Y-axes species are given in each panel. The markers are colour-coded by radio power, as given in the colour-bar. The orthogonal regression fit is also plotted, with 95\% confidence bands.}
 \label{fig:tspecdatapaper08}
\end{figure*}

We also correlate NIR emission line luminosities with each other; Figure \ref{fig:tspecdatapaper08} shows the plots of \pab vs. \HeI, \Fe{} and \Htwo, as well as \Fe{} vs. \Htwo. These all show strong correlation with each other, and display the trend of line strength vs. radio power, as color-coded for each marker. We note the particularly strong association between \pab{} and \HeI, especially at higher powers; this indicates that the excitation mechanisms for both species are very similar. The higher emission powers are almost certainly due to AGN activity; as these galaxies are early types, their SF rate is expected to be low. As an indication, a SF rate of 1 \msy{} will produce a \pab{} luminosity of the order of 10\pwr{40} \es{} \citep{Kennicutt1998}. Similarly, using the relationship in \cite{Brown2017} between radio power and \Ha{} luminosity, with $\log(P) = 21$, the SFR (without any AGN contribution) is 0.63 \msy.

\subsection{Aperture Effects and Emission Line Detection Rates}
The Triplespec spectrograph slit was set to 1\arcsec, and the data reduction pipeline used a 3\arcsec{} extraction aperture. At a distance of 1 Mpc, this scales to a window $\sim5 \times 15$ pc; our objects range in distance $7.4 - 167$ Mpc, i.e. window ranges of $37\times111$ pc -- $850\times2505$ pc. This could have an effect on detection, as the nuclear emission lines get diluted by increased galactic light with distance. As the integration time is mostly the same for all galaxies, further galaxies, being fainter in general, will decrease the S/N for each observation, again decreasing detectability.

\begin{figure*}
 \centering
 \includegraphics[width=.8\linewidth]{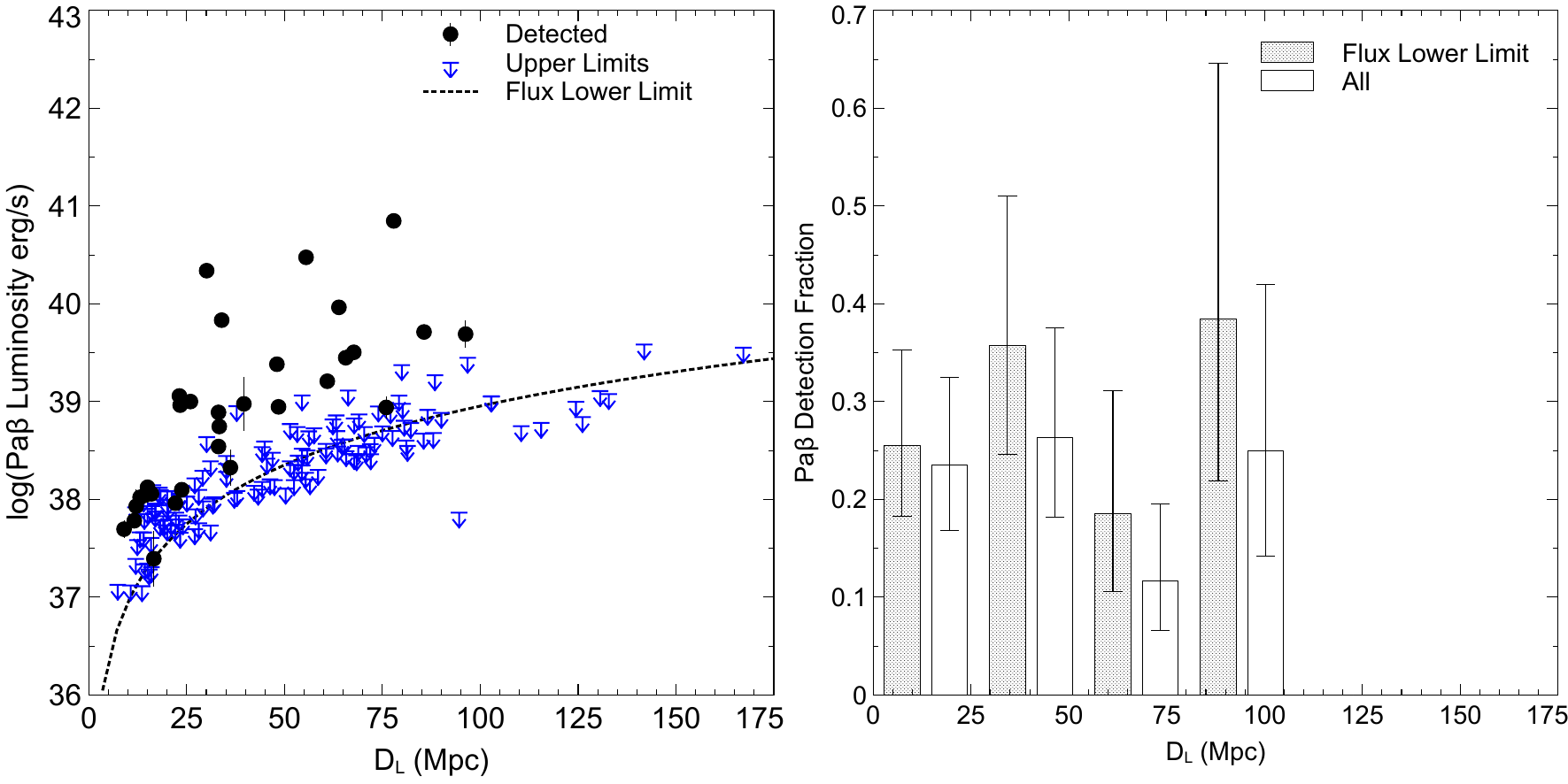}
 \caption{Left panel: \pab{} log luminosities and upper limits vs. luminosity distance. The flux lower limit of $7.5\times10^{-16}$ \ecs{} is plotted. Right panel: Histogram of \pab{} detection fraction vs. luminosity distance bin, showing fraction for all objects and for those above the flux lower limit. No bias of detection rates with luminosity distance is apparent.}
 \label{fig:tspecdatapaper12}
\end{figure*}

From the emission line measurements, the lowest flux detection for \pab{} is about 7.5~$\times$~10\pwr{-16}~\ecs. It is somewhat lower for \Htwo, which is due to the continuum having better S/N in that wavelength region. Plotting \pab{} luminosities and upper limits against luminosity distance (Figure \ref{fig:tspecdatapaper12} - left panel), we can see that there are no detections above 100 Mpc; this is beyond the nominal limit of 75 Mpc of \citetalias{Mould2012} We also plot the expected luminosity lower limit, given the flux detection limit. We also plot the histogram of detected vs. non-detected fractions of \pab{} at 25 Mpc intervals (Figure \ref{fig:tspecdatapaper12} - right panel), showing the fraction of detections for both all objects and those above the flux lower limit. The error bars are computed using Gehrel's approximation for small numbers \citep{Gehrels1986}; no definitive bias with distance is visible. The \pab{} detection rate for all objects is 19.6\%, if we exclude those objects below the flux detection threshold, this increases to 26.7\%.

If non-nuclear emission were significant, we would expect a bias on Figure \ref{fig:tspecdatapaper03} towards the left and down, i.e. for increasing SF. However, we see no such bias; our sample is largely confined to early-type galaxies, where we would not expect significant non-nuclear star formation. Similarly, the line index measurements (Figure \ref{fig:tspecdatapaper11}) show no bias with distance. 

\section{The Complete ``Radio Active'' Catalog}
\label{sec:TSpecRadioCatalog}
The complete catalog of radio galaxies (including the observed targets of ``opportunity'') has the object name and alternate ID (the primary identifier from SIMBAD or the NASA/IPAC Extragalactic Database (NED){\footnote{\url{http://ned.ipac.caltech.edu/}}} if not the same), the position and redshift, the $K$-band relative and absolute magnitude, the activity type and the morphology. The data in the original catalog has been extended to include the central velocity dispersion ($\sigma$) from the HyperLeda database\footnote{\url{http://leda.univ-lyon1.fr/}} \citep{Makarov2014} (440 objects) or from the SDSS spectroscopic data (11 objects) \citep{Abdurrouf2022,Smee2013}. 

The morphological types (T type) have been consolidated to the nearest integer (e.g. an original type of -3.3 is now set to -3). 
The SMBH mass (\mbh) is calculated from either $\sigma$ or from the \textit{K}-band magnitude (where $\sigma$ is not available). The formula for the relationship between $\sigma$ and \mbh{} is given by :
\begin{align}
 \log(\mmbh)& =\alpha + \beta~\log\left( \dfrac{\sigma}{200}\right)
\label{eqn:TSpecAtlas01}
\end{align}
where $\sigma$ is given in \kms{} and the parameters depend on the morphological type, as given by \cite{Sahu2019} (Table \ref{tbl:TSpecAtlas04}).

\begin{table}
 \centering
 \caption{Parameters for Equation \ref{eqn:TSpecAtlas01}.}
 \label{tbl:TSpecAtlas04}
\begin{tabular}{llrr}
\toprule
Morphology&T type &$\alpha$&$\beta$\\
\midrule
With a disk (ES, S0, Sp types)&T $>=$ -3& 5.72 & 8.22\\
Without a disk (E type)&  T $<=$ -4&6.69&8.65\\
\bottomrule
\end{tabular}
\end{table}
If $\sigma$ is not available, then \mbh{} is calculated from the formula presented by \cite{Graham2013} (from their Table 4), which is a broken power law:
\begin{align}
 \log(\mmbh) &= 9.05 -0.44\times (M_K + 25)& (M_K <-23.36)\\ 
&= 7.39 - 1.09\times (M_K + 22.5)&(M_K >=-23.36)
\end{align}

Radio data for the extended catalog is mostly from the NRAO VLA Sky Survey (NVSS) \citep{Condon1998} at 1.4 GHz, as described in \citetalias{Brown2011}, without any supplementary observations (525 objects in total). Where this was not available, data was taken from (in preference order) the VLA FIRST survey \citep[2 objects,][]{Becker1995} at 1.4 GHz, the Arecibo 2.38 GHz survey of bright galaxies \citep[5 objects,][]{Dressel1978}, the Sydney University Molonglo Sky Survey (SUMSS) at 843 MHz \cite[12 objects,]{Mauch2003}, the Parkes-MIT-NRAO catalog at 4.85 GHz\citep[1 object,][]{Gregory1994} and \cite[1 object,][]{Whiteoak1970} at 5 GHz. One object (NGC~~6831) had no radio flux information in the NASA/IPAC Extragalactic Database (NED). 

Distances were found from either the COSMICFLOWS-3 catalog \citep{Tully2016} or from the object redshift. The luminosity distance is computed from the distance using the functionality implemented in TOPCAT\footnote{\url{http://www.star.bristol.ac.uk/~mbt/topcat/}} \citep{Taylor2005}, which depends on the cosmological parameters given above, and is subsequently used to calculate NIR and radio luminosities. The 2MASS \citep{Skrutskie2006} \textit{K}-band absolute magnitude and distance are taken from the full table of data from \citetalias{Brown2011}.

Morphological (T) type is taken from the Third Reference Catalog of Bright Galaxies \citep{DeVaucouleurs1991}. Nuclear activity is taken from ``A catalogue of quasars and active nuclei: 12th edition'' \citep{Veron-Cetty2006} or, alternately, from SIMBAD (object type) or NED data (activity type).  These have been consolidated somewhat into the following values: Sy - Seyfert (not sub-categorised), Sy1/Sy1.5/Sy1.8/Sy1.9/Sy2 - Seyfert with sub-category, LIN - LINER, HII - H II, SBG - star-burst, AGN - AGN, QSO, blazar, BL Lac etc. (without further characterisation), Em - galaxies with emission lines but otherwise unclassified.

The catalog also lists other optical and NIR surveys for each object with either spectroscopy or IFU observations. These are listed in the ``Surveys'' column in Table \ref{tbl:TSpecAtlas05}, with counts of matched objects.

\begin{subtables}
\label{tbl:TSpecAtlas05}
\begin{table*}
\caption{Sample of the catalog of ``Radio Active'' galaxies, extending the \citetalias{Brown2011} catalog - Part 1. (The full table is available online)}
\label{tbl:TspecAtlas05a}
\centering
\begin{tabular}{@{}llcrrrrrrcrrc@{}}
\toprule
Galaxy                  & Alt. ID & Catalog & RA       & Dec      & z       & m$_K$ & M$_K$  & D     & Ref  & D$_L$ & $\sigma$ & Ref  \\
(1)                     & (2)     & (3)     & (4)      & (5)      & (6)     & (7)   & (8)    & (9)   & (10) & (11)  & (12)     & (13) \\ \midrule
2MASX J16251687-0910524 &         & 3       & 246.3204 & -9.1811  & 0.02861 & 10.20 & -25.20 & 117.6 &      & 120.1 &          &      \\
IC0450                  & Mrk 6   & 2       & 103.0514 & 74.4270  & 0.01302 & 9.56  & -24.90 & 76.8  &      & 77.9  &          &      \\
IC5063                  &         & 2       & 313.0097 & -57.0688 & 0.01127 & 8.75  & -24.59 & 46.3  &      & 46.6  & 152.3    &      \\
NGC0080                 &         & 1       & 5.2951   & 22.3571  & 0.01895 & 8.92  & -25.71 & 83.2  & C    & 84.5  & 248.4    &      \\
NGC0447                 &         & 2       & 18.9069  & 33.0678  & 0.01885 & 9.48  & -25.00 & 77.5  &      & 78.6  & 157.8    &      \\
NGC1052                 &         & 1       & 40.2700  & -8.2558  & 0.00493 & 7.45  & -23.94 & 19.0  &      & 19.0  & 208.0    &      \\
NGC1316                 &         & 1       & 50.6741  & -37.2082 & 0.00591 & 5.59  & -26.03 & 21.0  & C    & 21.0  & 223.1    &      \\
NGC2217                 &         & 1       & 95.4156  & -27.2338 & 0.00543 & 7.09  & -24.94 & 25.5  & C    & 25.5  & 216.1    &      \\
NGC3182                 &         & 2       & 154.8876 & 58.2057  & 0.00706 & 9.47  & -22.84 & 29.0  &      & 29.1  & 112.7    & S    \\
NGC3593                 &         & 2       & 168.6541 & 12.8182  & 0.00211 & 7.42  & -22.35 & 9.0   & C    & 9.0   & 73.9     &      \\
NGC4278                 &         & 1       & 185.0284 & 29.2808  & 0.00206 & 7.18  & -23.84 & 16.0  &      & 16.0  & 237.3    &      \\
NGC4438                 &         & 2       & 186.9403 & 13.0088  & 0.00035 & 7.27  & -23.06 & 11.6  & C    & 11.6  & 135.3    &      \\
NGC4486                 & M 87    & 1       & 187.7059 & 12.3911  & 0.00420 & 5.81  & -25.34 & 17.0  & C    & 17.0  & 323.0    &      \\
NGC4694                 &         & 1       & 192.0629 & 10.9836  & 0.00388 & 8.96  & -22.06 & 16.0  &      & 16.0  & 55.5     &      \\
UGC3426                 & Mrk 3   & 1       & 93.9012  & 71.0375  & 0.01300 & 8.97  & -24.75 & 55.0  &      & 55.5  & 244.6    &    \\ \bottomrule
\end{tabular}
\begin{flushleft}
\small
\begin{tabular}{ll}
\textbf{Notes:}\\
1&Galaxy identifier\\
2&Alternative identifier.\\
3&Catalog: 1 - \citetalias{Brown2011}, 2 - extended, 3 - supplementary.\\
4,5,6&RA, Dec and redshift $z$ from are from SIMBAD.\\
7,8&$m_K$ from the 2MASS catalog \citep{Skrutskie2006}, $M_K$ calculated from the luminosity distance.\\
9&Distance (Mpc) from COSMICFLOWS-3 or NED, as described in the text.\\
10&C - source of distance information from COSMICFLOWS-3.\\
11&Luminosity distance calculated from TOPCAT functionality.\\
11&Stellar velocity dispersion in \kms.\\
12&Velocity dispersion source; no flag - HyperLeda, S - SDSS spectra.\\
\end{tabular}
\end{flushleft}
\end{table*}
\begin{table*}
\caption{Sample of the catalog of ``Radio Active'' galaxies, extending the \citetalias{Brown2011} catalog - Part 2. (The full table is available online)}
\label{tbl:TspecAtlas05b}
\centering
\begin{tabular}{@{}lclcrcrrccrcl@{}}
\toprule
Galaxy                  & T Type & Activity & Mod.  & log \mbh & Ref  & Flux     & e\_Flux & Ref  & UL   & log P & Obs  & Surveys    \\
(1)                     & (14)   & (15)     & (16) & (17)       & (18) & (19)     & (20)    & (21) & (22) & (23)  & (24) & (25)       \\ \midrule
2MASX J16251687-0910524 & -1     &          &      & 9.14       & K    & 4.6      & 0.5     &      &      & 21.90 & *    & 17         \\
IC0450                  & -0.5   & Sy1.5    &      & 9.00       & K    & 269.5    & 8.1     &      &      & 23.29 & *    & 14         \\
IC5063                  & -0.8   & Sy1      &      & 7.54       &      & 1975.0   & 59.3    & S    &      & 23.71 &      & 12 14 17   \\
NGC0080                 & -2.5   &          &      & 8.76       &      & 0.9      & 0.5     &      & <    & 20.89 &      &            \\
NGC0447                 & 0      &          &      & 7.63       &      & 6.0      & 4.0     & A    & <    & 21.65 &      & 3          \\
NGC1052                 & -5     & LIN      &      & 8.76       &      & 1100.0   & 100.0   &      &      & 22.68 & *    & 7 10 11 12 \\
NGC1316                 & -2     & AGN      &      & 8.49       &      & 150000.0 & 10000.0 &      &      & 24.90 &      & 10         \\
NGC2217                 & -1     & LIN      &      & 8.41       &      & 22.0     & 1.0     &      &      & 21.24 &      & 10 12      \\
NGC3182                 & 1      & Sy2      &      & 6.80       &      & 2.3      & 0.5     &      &      & 20.37 & *    & 1 3 13 17  \\
NGC3593                 & 0      & Sy2      & *    & 5.75       &      & 86.2     & 3.4     &      &      & 20.92 & *    & 8 13 17    \\
NGC4278                 & -5     & LIN      &      & 9.15       &      & 390.0    & 10.0    &      &      & 22.08 & *    & 11 13      \\
NGC4438                 & 0      & LIN      &      & 7.25       &      & 63.3     & 2.7     &      &      & 21.01 & *    & 11         \\
NGC4486                 & -4     & LIN      &      & 10.04      &      & 210000.0 & 10000.0 &      &      & 24.86 &      & 1 11 13    \\
NGC4694                 & -2     & HII      &      & 5.04       &      & 3.5      & 0.5     &      &      & 20.03 & *    & 1 13 16    \\
UGC3426                 & -2     & Sy1      &      & 8.72       &      & 1100.0   & 100.0   &      &      & 23.61 & *    &     \\ \bottomrule    
\end{tabular}
\begin{flushleft}
\small
\begin{tabular}{ll}
\textbf{Notes:}\\
1&Galaxy identifier\\
14&Morphological type from \cite{DeVaucouleurs1991}. \\
15&Nuclear activity type, consolidated from \citep{Veron-Cetty2006}, SIMBAD or NED.\\
16&Activity type modified, * - re-measured from NED spectra (see text).\\
17&Black hole mass (units of log \msun), computed as described in text.\\
18&\mbh{} source, K -  computed from M$_K$, otherwise from $\sigma$.\\
19,20&Radio flux and uncertainty (mJ).\\
21&Radio flux reference. blank - NVSS, F - VLA FIRST, A - Arecibo, S - SUMSS, P - Parkes-MIT-NRAO, W- Whiteoak.\\ 
22&Radio upper limit flag.\\
23&Log radio power (W Hz\pwr{-1}).\\
24&Observed. * - observed with TripleSpec. \\
25&Surveys. As described in the main text and listed in Table \ref{tbl:TSpecAtlas06}.
\end{tabular}
\end{flushleft}
\end{table*}
\end{subtables}
\begin{table}
    \caption{Spectral and IFU surveys, for Table \ref{tbl:TSpecAtlas05}}
    \label{tbl:TSpecAtlas06}
    \centering
\begin{tabular}{clrl}
\toprule
ID& Survey & \#& Reference\\\midrule
1  & ATLAS3D-SAURON & 36  & \cite{Cappellari2011}       \\
2  & SAMI           & 2   & \cite{Scott2018}            \\
3  & CALIFA         & 2   & \cite{Sanchez2012}          \\
4  & SDSS MANGA     & 2   & \cite{Abdurrouf2022}        \\
5  & KONA           & 1   & \cite{Muller-Sanchez2018a}  \\
6  & LLAMA          & 1   & \cite{Lin2018}              \\
7  & AGN-IFS        & 3   & \cite{Storchi-Bergmann2013} \\
8  & NUGA           & 4   & \cite{Garcia-Burillo2007}   \\
9  & GATOS          & 1   & \cite{Garcia-Burillo2021}   \\
10 & DIVING3D       & 44  & \cite{Steiner2022}          \\
11 & PALOMAR        & 21  & \cite{Filippenko1985}       \\
12 & 6dFGS          & 246 & \cite{Jones2004,Jones2009}  \\
13 & SDSS Spectra   & 77  & \cite{Abdurrouf2022}        \\
14 & BASS           & 22  & \cite{Koss2022}             \\
15 & GAMA           & 7   & \cite{Driver2009}           \\
16 & PHANGS         & 6   & \cite{Leroy2021}            \\
17 & AllskyAGN      & 76  & \cite{Zaw2019}              \\
\bottomrule
\end{tabular}
\end{table}
\section{Conclusions}
We have presented a near infrared spectroscopic atlas of nearby, bright early-type galaxies with radio emission, containing 163 galaxies observed by the Palomar 200\arcsec{} TripleSpec instrument. We explore the science goals enabled by these observations. We have measured the emission line fluxes for H, He, \Fe{} and \Htwo{} and determined the nuclear excitation mechanisms, noting a population  which are optically classified as star-forming, but have AGN emission line ratios in the infrared, most likely due to the better dust penetration in the infrared. 

By stacking spectra showing \Htwo{} emission, we deduced the \Htwo{} excitation temperature ($1957\pm182$ K) and dominant excitation mechanism (thermal and shock heating combined) from the \textit{K}-band emission line sequence. When binned by low or high power radio emission, the lower power bin is warmer than the higher power. This may be explained by the relative balance of heating mechanisms changing at higher radio power.

By stacking those spectra that have no emission lines, we have produced an ``average'' spectrum of the continuum, showing the main absorption features and the proposed spectral indices from the literature. Plotting the CO12 absorption line index vs. \jk{} colour shows a trend, with stronger nuclear activity producing a weaker CO12 index and a redder (flatter) continuum.

We have explored the correlations between the radio and various emission-line luminosities; finding a trend with radio power for all NIR emission lines. However, we deduced from the large scatter in the upper limits to NIR emission lines that the two are not directly coupled and that the duty cycles of SF and AGN activity are not synchronised.

We have also presented the complete early-type radio galaxy catalog, where the original published in \citetalias{Brown2011}, which focused on elliptical and S0 galaxies, has been supplemented with S0/a plus some later type galaxies, with supplementary objects that were observed in the NIR with Triplespec, making a total of 546 objects. This catalog contains \textit{K} band magnitudes, stellar dispersions, distances, black hole mass estimates and radio flux and power, as well as morphology and activity information.
\section*{Acknowledgements}
We would like to thank our Triplespec colleagues at Caltech
for their support of our survey; thanks go to Eilat Glickman for sharing her observing expertise at the Palomar 200\arcsec{} telescope and her copy of the Triplespec SpexTool IDL program. The Caltech-Swinburne Collaborative Agreement allowed JM access to the Hale telescope. This research has made use of the NASA/IPAC Extragalactic Database (NED) which is operated by the Jet Propulsion Laboratory, California Institute of Technology, under contract with the National Aeronautics and Space Administration. This publication makes use of data products from the Two Micron All Sky Survey, which is a joint project of the University of Massachusetts and the Infrared Processing and Analysis Center/Caltech, funded by NASA and the U.S. National Science Foundation. This research has also made use of the SIMBAD database, operated at CDS, Strasbourg, France \citep{Wenger2000}.

This paper makes use of SDSS data. Funding for the Sloan Digital Sky Survey IV has been provided by the Alfred P. Sloan Foundation, the U.S. Department of Energy Office of Science, and the Participating Institutions. SDSS acknowledges support and resources from the Center for High-Performance Computing at the University of Utah. The SDSS web site is www.sdss.org.

SDSS is managed by the Astrophysical Research Consortium for the Participating Institutions of the SDSS Collaboration including the Brazilian Participation Group, the Carnegie Institution for Science, Carnegie Mellon University, Center for Astrophysics, Harvard \& Smithsonian (CfA), the Chilean Participation Group, the French Participation Group, Instituto de Astrofísica de Canarias, The Johns Hopkins University, Kavli Institute for the Physics and Mathematics of the Universe (IPMU) / University of Tokyo, the Korean Participation Group, Lawrence Berkeley National Laboratory, Leibniz Institut f{\"u}r Astrophysik Potsdam (AIP), Max-Planck-Institut f{\"u}r Astronomie (MPIA Heidelberg), Max-Planck-Institut f{\"u}r Astrophysik (MPA Garching), Max-Planck-Institut f{\"u}r Extraterrestrische Physik (MPE), National Astronomical Observatories of China, New Mexico State University, New York University, University of Notre Dame, Observat{\'o}rio Nacional / MCTI, The Ohio State University, Pennsylvania State University, Shanghai Astronomical Observatory, United Kingdom Participation Group, Universidad Nacional Aut{\'o}noma de México, University of Arizona, University of Colorado Boulder, University of Oxford, University of Portsmouth, University of Utah, University of Virginia, University of Washington, University of Wisconsin, Vanderbilt University, and Yale University.

\textit{Facilities:} Triplespec on Palomar 200\arcsec{} 
\section*{Data Availability}
Tables \ref{tbl:TSpecAtlas02}, \ref{tbl:TSpecAtlas03},  \ref{tbl:TSpecAtlas05} and \ref{tbl:TSpecAtlas06} plus the reduced, restframe spectra are available in full at CDS via anonymous ftp to \url{cdsarc.u-strasbg.fr} (\url{130.79.128.5}) or via \url{https://cdsarc.unistra.fr/viz-bin/cat/J/MNRAS/???/???}.
\bibliographystyle{mnras}
\bibliography{library} 
\bsp 
\label{lastpage}
\end{document}